\documentclass[onecolumn,aip,jcp,preprint,amsmath,amssymb,superscriptaddress,showpacs]{revtex4-1}

\usepackage[pdftex]{graphicx}
\usepackage{graphics}
\usepackage{dcolumn}
\usepackage{bm}
\usepackage{color}
\usepackage{soul}
\usepackage{textcomp}
\usepackage{soul}
\usepackage{rotating}
\usepackage{setspace}
\usepackage[mathlines]{lineno}
\usepackage{textcomp}
\usepackage{mathcomp}

\def\bra#1{\langle\,#1\,|}
\def\ket#1{|\,#1\, \rangle}

\usepackage{microtype}
\usepackage[breaklinks=true,hyperindex=true,
pdftitle={Mechanisms in Environmentally-Assisted One-photon  Phase Control},
colorlinks=true,pagebackref=false,citecolor=blue,plainpages=false,pdfpagelabels,
linkcolor=blue,urlcolor=blue]{hyperref}

\def\iu{{\rm i}}

\begin{document}

\title{
Mechanisms in Environmentally-Assisted One-photon  Phase Control
}

\author{Leonardo A. Pach\'on}
\affiliation{Chemical Physics Theory Group, Department of Chemistry and
Center for Quantum Information and Quantum Control,
\\ University of Toronto, Toronto, Canada M5S 3H6}
\affiliation{Grupo de F\'isica At\'omica y Molecular, Instituto de F\'{\i}sica,  Facultad de Ciencias Exactas y Naturales, 
Universidad de Antioquia UdeA; Calle 70 No. 52-21, Medell\'in, Colombia}
\author{Paul Brumer}
\affiliation{Chemical Physics Theory Group, Department of Chemistry and
Center for Quantum Information and Quantum Control,
\\ University of Toronto, Toronto, Canada M5S 3H6}

\begin{abstract}
The ability of an environment to assist in one-photon phase control relies
upon entanglement between the system and bath and on the
breaking of the time reversal symmetry.
Here, one photon phase control is examined analytically and numerically
in a model system, allowing an analysis of
the relative strength of these contributions. Further,
the significant role of non-Markovian dynamics and of
moderate system-bath coupling in enhancing one-photon phase control
is demonstrated, and an explicit role for quantum mechanics
is noted in the existence of initial non-zero stationary coherences.
Finally, desirable conditions are shown to be required to
observe such environmentally assisted control, since 
the system will naturally equilibrate with its environment at longer times, ultimately 
resulting in the loss of phase control.

\end{abstract}

\date{\today}

\maketitle

\def\linenumberfont{\normalfont\tiny}
\section{Introduction}
The coherent control of molecular processes as been highly successful, both computationally
and experimentally, when applied to isolated molecular systems \cite{SB12}.
Indeed, a wide variety of scenarios has been proposed ranging from the control of
products in the continuum, such as photodissociation, to bound state control of
radiationless transitions\cite{SH11,SH12,GSB13} . 
By contrast, control of a system in the presence of an environment, where decoherence
effects often destroy coherences\cite{Sch07,EB13}.
is only in its early stages of development. To this end, one photon phase control, the
subject of this paper, is of particular interest insofar as the environment \textit{assists},
rather than impedes,  
control of the system dynamics.

One photon phase control has been the subject of considerable recent discussion and attention
\cite{PM&06,Jof07,PM&07,FC&09,KRK10,SAB10,KR&11,PH&11,faraday153,AB13,PYB13}. In this
one photon phase control (OPPC) scenario, control is achieved by varying the spectral
phase of a weak pulse while keeping its power spectrum fixed. Particular interest
in this scenario arises from a seminal proof in which it was shown that OPPC was 
not possible for isolated molecular systems in which control was over products in the continuum\cite{BS89}.
However, subsequent experiments on control of retinal isomerization in bacteriorhodopsin in the 
weak regime\cite{PM&06} as well as in the strong field regime\cite{FC&09} motivated 
controversy\cite{Jof07,PM&07,FC&09,PH&11} and the need for clarification of conditions 
under which such control was possible.

This clarification, provided in Refs.~\citenum{SAB10} and \citenum{PYB13}, showed that 
control was {\it possible} for both isolated systems and open quantum systems under 
well defined conditions.
In particular, for an observable $\hat{O}$ of a physicochemical system defined by the Hamiltonian 
$\hat{H}_{\mathrm S}$, it was shown that:
(i) if the system is isolated and initially devoid of coherence then one-photon phase control is possible only if 
$[\hat{H}_{\mathrm S},\hat{O}] \ne 0$,
but 
(ii) if the system is coupled to an environment, as in any realistic case, then control is 
possible not only if $[\hat{H}_{\mathrm S},\hat{O}] \ne 0$, but even if $[\hat{H}_{\mathrm S},\hat{O}] = 0$, 
in which case it is \textit{environmentally assisted}\cite{SAB10,AB13,PYB13}.
As it was noted in Refs.~\citenum{SAB10} and \citenum{PYB13}, the case of
$[\hat{H}_{\mathrm S},\hat{O}] \ne 0$ includes, e.g., isomerization since
the probability of 
observing an isomer is an observable that does not commute 
with $\hat{H}_{\mathrm S}$.

The aim of Ref.~\citenum{SAB10} was to establish these general commutation-based rules, 
whereas Ref.~\citenum{PYB13} identified the physical processes responsible for OPPC.
In particular, based on a general master equation approach, we qualitatively identified two main 
mechanisms for OPPC in open systems:
the breaking of time-reversal symmetry and the entanglement between 
the system and the bath.
These two mechanisms differ in character; the time-reversal symmetry does not 
rely upon quantum mechanics, 
whereas the initial correlations between the system and the bath are quantum in nature\cite{PTB13}.
These mechanisms also 
work on different time scales, the initial correlations contribute in the 
short time regime while the time-reversal symmetry dominates in the long time regime.
As such, it is the latter that determines the amount of control that can be achieved within this control
scheme.

In this paper, we quantitative analyze the magnitude of each of these contributions and 
demonstrate 
the significant role of the non-Markovian dynamics and of moderate system-bath coupling 
in enhancing the extent of one-photon phase control.
Note that the system-bath interaction plays a dual role in the OPPC scenario. First,
it assists insofar as allowing phase control for systems where such control would
not occur if the system were isolated. However, since the system-bath coupling
persists long after the laser pulse, it induces relaxation to equilibrium at long
times, resulting in long-time loss of phase control. Hence, as we show below,
maintaining phase control over an extended period of time requires careful balancing
of the system-bath interactions.


\section{Initial considerations}
Consider a quantum system S described by the Hamiltonian
$\hat{H}_{\mathrm{S}}$ with $\hat{H}_{\mathrm{S}}\ket{n} = E_n \ket{n}$.  
We consider below two cases, where the system is isolated and irradiated with a laser, and 
the second where the system is irradiated in the presence of an environment (or ``bath").  
In the latter case, the full Hamiltonian is given by 
$\hat{H} = \hat{H}_{\mathrm S} + \hat{V}_{\mathrm{L}}+\hat{H}_{\mathrm B} + \hat{H}_{\mathrm{SB}}$, 
where $\hat{V}_{\mathrm L}$ denotes the term laser, $\hat{H}_{\mathrm B}$ is the Hamiltonian of the 
environment and $\hat{H}_{\mathrm{SB}}$ is the system-bath coupling.

\subsection{Initial Considerations on System Dynamics}
Under the influence of time-dependent fields
and in the presence of an environment, the time evolution of the system density-operator
$\hat{\rho}_{\mathrm{S}}$ is given by \cite{PB13}
\begin{align}
\label{equ:RedDenMatEinBas}
\begin{split}
\langle n | \hat{\rho}_{\mathrm{S}}(t) | m \rangle &=
\sum_{\nu} J_{nm;\nu\nu}(t)
\langle \nu | \hat{\rho}_{\mathrm{S}}(0) | \nu \rangle
\\&+
\sum_{\nu\neq\mu} J_{nm;\nu\mu}(t)
\langle \nu | \hat{\rho}_{\mathrm{S}}(0) | \mu \rangle,
\end{split}
\end{align}
where $J_{nm;\nu\mu}(t)$ denotes the propagation function in the system energy basis
representation \cite{PB13}. As such,
the system dynamics is contained in the propagation function
elements $J_{nm;\nu\mu}(t)$ that can be derived, as in Ref.~\citenum{PB13}, from a path
integral representation of the propagation function.
The first two indices $nm$ denote the density matrix element in which we are interested and 
the last two indices $\nu\mu$ refer to the elements of the initial density matrix that contribute, 
as shown in Eq.~(\ref{equ:RedDenMatEinBas}), to the dynamics of the $nm$-th element.
The general picture implied by Eq.~(\ref{equ:RedDenMatEinBas}) is that the time
evolution of the diagonal elements $n=m$ (populations) as well as the off-diagonal
elements (coherences) of the density matrix
depends linearly on the initial diagonal and off-diagonal elements.

To explore the physical meaning and role of the $J_{nm;\nu\mu}(t)$
consider first the case of unitary time-evolution in
absence of time dependent external forces. 
Examining this case allows a reinterpretation of some of the known features of 
unitary dynamics in terms of this formalism.
In the case of unitary time-evolution, the elements of the propagating function in
Eq.~(\ref{equ:RedDenMatEinBas}) reduce to the familiar expression
\begin{equation}
\label{PropFuncElemUniEvo}
J_{nm;\nu\mu}(t) = \mathrm{e}^{-\mathrm{i}(E_{m} - E_{n})t/\hbar}
\delta_{n\nu}\delta_{m\mu}.
\end{equation}
The  Kronecker deltas prevent
(a) the transfer of initial population from $\bra{\nu}\hat{\rho}_{\mathrm{S}}(0)\ket{\nu}$ to
$\bra{n}\hat{\rho}_{\mathrm{S}}(t)\ket{n}$ mediated by $J_{nn;\nu\nu}(t)$,   as well as (b) the generation
of coherences, $n\neq m$, from initial populations $\langle\nu|\hat{\rho}_{\mathrm{S}}(0)|\nu\rangle$.
In addition, the possibility of controlling the populations at time $t$, i.e., the $\langle n|\hat{\rho}_{\mathrm{S}}(t)|n\rangle$, 
by varying the initial coherences $\langle \nu|\hat{\rho}_{\mathrm{S}}(0)|\mu\rangle$, $\nu \ne \mu$,
a possible control objective, is also prevented
during the unitary time evolution governed by Eq.~(\ref{PropFuncElemUniEvo}).

In the presence of time dependent external fields or of an external environment, the propagating
function elements $J_{nm;\nu\mu}(t)$ differ from those in Eq.~(\ref{PropFuncElemUniEvo}), as discussed 
below.
Although the particular form of the $J_{nm;\nu\mu}(t)$ depends upon the
environment and on the external field, generally the Kronecker
delta restrictions will disappear, allowing for the transfer of population, the generation
of coherences and the contribution of the initial coherences to the time
evolution of the populations.
Specifically, in the presence of dissipation, the delta functions in Eq.~(\ref{PropFuncElemUniEvo})
broaden, allowing for the additional processes  discussed below. 
In Ref.~\citenum{PB13} we provide
a deeper description of these extra processes in the context of incoherent
excitation of open quantum systems \cite{JB91,HB11,BS12b,PB12b,PB13}.
These effects are quantitatively considered, for OPPC, below.

\subsection{Initial Considerations on Control}

Consider now controlling the expectation value of a system observable $\hat{O}$. 
In general, the time evolution of the
expectation value  $\hat{O}$ can be expressed as
$
\langle \hat{O}(t) \rangle = \sum_{n,m}
\langle n | \hat{O} | m \rangle \langle m|\hat{\rho}_{\mathrm{S}}(t)|n\rangle.
$
We are particularly interested in observables where $[\hat{O},\hat{H}_{\mathrm{S}}] = 0$,
i.e. those that 
are not phase controllable \cite{SAB10,PYB13} in isolated systems.
In this case,
$
\langle \hat{O}(t) \rangle = \sum_{n}
\langle n | \hat{O} | n \rangle \langle n|\rho_{\mathrm{S}}(t)|n\rangle.\,
$
Using Eq.~(\ref{equ:RedDenMatEinBas}), we obtain
\begin{align}
\label{ExpValuObsProFun}
\begin{split}
\langle \hat{O}(t) \rangle &= \sum_{n,\nu}
\langle n | \hat{O} | n \rangle  J_{nn;\nu\nu}(t)
\langle \nu |\hat{\rho}_{\mathrm{S}}(0) | \nu \rangle
\\&+
\sum_{n,\nu\neq\mu} \langle n | \hat{O} | n \rangle J_{nn;\nu\mu}(t)
\langle \nu |\hat{\rho}_{\mathrm{S}}(0) | \mu \rangle.
\end{split}
\end{align}

First consider one-photon phase control from states initially devoid 
of coherence, i.e., $\langle \nu|\hat{\rho}_{\mathrm{S}}(0)|\mu \rangle = 0$ 
for $\nu \ne \mu$.
Under these circumstances, and with the assumption of irradiation with
\emph{weak} laser fields, one-photon phase control is not possible  if 
the molecule is isolated, i.e., if the dynamics is
unitary\cite{SAB10,PYB13}. 
By contrast, for the case of non-unitary dynamics (the open system case), 
environmentally assisted control is possible\cite{SAB10,PYB13}.

Below, we examine the origin and nature of this effect.  
Although Eq.~(\ref{ExpValuObsProFun}) is general, we anticipate two
physical mechanisms that could be responsible for one-photon
phase control in the open-system case.
The first arises from the fact that, in the presence of the environment,
the stationary states 
of the system are no longer diagonal in the system's Hamiltonian eigenbasis $\{\ket{n}\}$. 
Specifically, interaction with  the environment  produces off-diagonal elements in the 
initial system density operator $\hat{\rho}_{\mathrm{S}}(0)$ which, by virtue of Eq.~(\ref{ExpValuObsProFun}), will
contribute to the evolution of the expectation value of $\hat{O}$.
The second relates to time-reversal-symmetry breaking. 
For unitary evolution,  one can decompose $J_{nn;\nu\nu}(t)$
into two terms, one related to the process $\ket{\nu}\rightarrow\ket{n}$
and another term related to the dual process, $\bra{n}\leftarrow\bra{\nu}$.
These two processes ``interfere destructively" in $J_{nn;\nu\nu}(t)$. However, in the particular
case of an open system this symmetry is broken, allowing for the encoding
of phase information in $J_{nn;\nu\nu}(t)$, as noted below.  

A  detailed  analysis  of  phase  control  in  an isolated and
model system follows below.

\section{Unitary Evolution and One-photon Phase-control}
\label{sec:UnitEvouOPPC}
To examine one-photon phase control we consider an analytically soluble model for both  
the unitary and non-unitary cases.
In particular, consider the vibrations of a diatomic molecule of frequency
$\omega_0 = \sqrt{k/m}$, where $m$ is the reduced mass and $k$ the coupling
constant between the atoms. 
Although an apparently simple model, it will be seen to provide great insight into the physics 
of  phase  control.  It  also  provides  a  model  for  a wide variety of physical systems  
such as nano-mechanical resonators, trapped ions, membranes, optical mirrors, etc. \cite{GPZ10}.

For the unitary case, the Hamiltonian is given by
\begin{align}
\label{equ:HS}
\begin{split}
\hat{H} &= \hat{H}_{\mathrm{S}} + \hat{V}_{\mathrm{L}} =
\frac{\hat{p}^2}{2m} + \frac{m \omega_0^2}{2}\hat{q}^2 + \hat{q}E(t),
\end{split}
\end{align}
where $E(t)$ denotes the electric field of the laser pulse.
This Hamiltonian describes then a linearly-forced harmonic oscillator.
The Fourier transform of the electric field is  
$\tilde{E}(\omega) = \sqrt{|S(\omega)|} \exp[-\iu \varphi(\omega)]$, where $S(\omega)$ is the
field amplitude and $\varphi(\omega)$ is the spectral phase. 
``Phase control''  refers to the effects on the molecular dynamics  of manipulating
the spectral phase $\varphi(\omega)$, while keeping $S(\omega)$ fixed.

\subsection{System Unitary Time Evolution}

In the position representation, the time evolution of the
system-density-matrix element
$\rho_{\mathrm{S}}(\mathbf{q}',0)= \bra{q_+'}\hat{\rho}_{\mathrm{S}}(0)\ket{q_-'}$ can derived from
\begin{align}
\begin{split}
\rho_{\mathrm{S}}(\mathbf{q}'',t) =
\int \mathrm{d} q' J(\mathbf{q}'',t;\mathbf{q}',0)
\rho_{\mathrm{S}}(\mathbf{q}',0),
\end{split}
\end{align}
where $J(\mathbf{q}'',t;\mathbf{q}',0)$ is the propagating function, which
for the case of unitary time-evolution is
$J(\mathbf{q}'',t;\mathbf{q}',0)= U(q_+'',q_+',t) U^*(q_-'',q_-',t)$, with
$U(q_+'',q_+',t)  = \bra{q_+''}\hat{U}(t)\ket{q_+'}$ and
$\hat{U}(t) = \hat{\mathcal{T}}\exp[-\mathrm{i}\int_{t_0}^{t}\mathrm{d}s \hat{H}(s)/\hbar]$,
$\hat{\mathcal{T}}$ being the time-ordering operator
 \cite{FH65,Ing02,PID10}.
For the unitary case, with time evolution operator $\hat{U}(t)$, one can obtain
the propagating function elements $J_{nm;\nu\mu}(t)$ in Eq.~(\ref{equ:RedDenMatEinBas}) by
projecting onto the system energy basis \cite{PB13}.
For the particular case  in Eq.~(\ref{equ:HS}),
the unitary time evolution can be obtained analytically (c.f. Chap.~3 in
Ref.~\citenum{FH65} or Chap.~2 in Ref.~\citenum{Ing02}), giving analytic  propagating 
function elements.

For the purpose of  discussion,  consider the case where  the system is
prepared in a coherent superposition of the ground and first excited
states, i.e.,
$\hat{\rho}_{\mathrm{S}}(0) = \left(
\ket{0}\bra{0} + \ket{0}\bra{1} + \ket{1}\bra{0} + \ket{1}\bra{1}
\right)/2$.
From Eq.~(\ref{ExpValuObsProFun}), we have
\begin{equation}
\langle \hat{O}(t) \rangle = \frac{1}{2}
\sum_{\substack{n\\ \nu,\mu=0,1}}\langle n | \hat{O} | n \rangle
J_{nn;\nu\mu}(t),
\end{equation}
where
\begin{align}
\label{Equ:DefPropUnitEvoJnn00}
\begin{split}
J_{nn;00}(t)& = \frac{1}{n!\mathcal{E}_0^{2n}}
|\mathcal{E}(t)|^{2n}
\exp\left\{ -\frac{|\mathcal{E}(t)|^2}{\mathcal{E}_0^2 }\right\},
\end{split}
\\
\label{Equ:DefPropUnitEvoJnn01}
\begin{split}
J_{nn;01}(t)& =
\frac{1}{n!{\mathcal{E}_0}^{2n+1}}
|\mathcal{E}(t)|^{2n-2}
\exp\left\{ -\frac{|\mathcal{E}(t)|^2}{\mathcal{E}_0^2 }\right\}\\
&\times
\mathcal{E}(t)
\left\{-2n\mathcal{E}_0^2 + |\mathcal{E}(t)|^2\right\},
\end{split}
\\
\label{Equ:DefPropUnitEvoJnn11}
\begin{split}
J_{nn;11}(t)& =
\frac{1}{n!{\mathcal{E}_0}^{2n+2}}
|\mathcal{E}(t)|^{2n-2}
\exp\left\{ -\frac{|\mathcal{E}(t)|^2}{\mathcal{E}_0^2 }\right\}\\
&\times
\left\{n^2 \mathcal{E}_0^4 +  |\mathcal{E}(t)|^2
\left[ -2n\mathcal{E}_0^2 + |\mathcal{E}(t)|^2 \right]\right\},
\end{split}
\end{align}
$J_{nn;10}(t) = J_{nn;01}^*(t)$, $\mathcal{E}_0= \sqrt{2 m \omega_0 \hbar}$, and
\begin{align}
\label{equ:GenFouTraField}
\mathcal{E}(t) &= E_-^2(t) +2 \cos(\omega_0 t) E_-(t) E_+(t) + E_+^2(t)
\\
\label{equ:FouTraField}
&= \int_{-\infty}^{t} \mathrm{d}s \exp(-\iu s \omega_0) E(s).
\end{align}
For later convenience, we have defined
$
E_+(t) = \int_{-\infty}^t  \mathrm{d}s E(s) G_+(t-s)/G_+(t),
$ and  $
E_-(t) = \int_{-\infty}^t \mathrm{d}s E(s) G_-(s)/G_-(t),
$
where $G_{\pm}(s)$ denotes the inverse Laplace transform of
$\hat{G}_{\pm}(z) = \left(z^2 + \omega_0^2 \right)^{-1}$.
The specific combination of $E_+(t)$ and $E_-(t)$ in Eq.~(\ref{equ:GenFouTraField})
is responsible for the simple result in Eq.~(\ref{equ:FouTraField}).
As shown below, this precise form is lost  in the open system case giving 
rise to phase control.
In the following, we use linearly chirped laser pulses
\begin{equation}
\label{equ:pulse}
E(t) = E_0 \mathrm{e}^{-4 \left(\frac{t-t_0}{\Delta t} \right)^2}
\cos\left[ \Omega_{\mathrm{L}}(t-t_0) +\chi(t-t_0)^2 \right],
\end{equation}
where $t_0$ and $\Delta t$ are the center and width of the Gaussian pulse
envelope, $E_0$ and $\Omega_{\mathrm{L}}$ are the field strength
and carrier frequency of the laser with chirp rate  $\chi$.
Changing the sign of $\chi$ causes a change in the laser phase while 
retaining the intensity profile.  

In the long time regime, $t\rightarrow\infty$, $\mathcal{E}(t)$ becomes
the Fourier transform of the field $E(t)$.
This implies that the term $ |\mathcal{E}(t)|$ does not contain any information
about the spectral phase. Hence, here, laser phase information is not encoded 
in either $J_{nn;00}(t)$ or $J_{nn;11}(t)$, but only in $J_{nn;01}(t)$.
Thus, if the initial state is an incoherent superposition of 
$\ket{0}$ and $\ket{1}$ then the expectation value of 
$\hat{O}$ does not depend on the phase function $\varphi(\omega)$.
This result makes no reference to the field strength, but is a
particularity of the chosen system which is not expected to be  
true in general.

\begin{figure}[h]
\begin{tabular}{cc}
\includegraphics[width = \columnwidth]{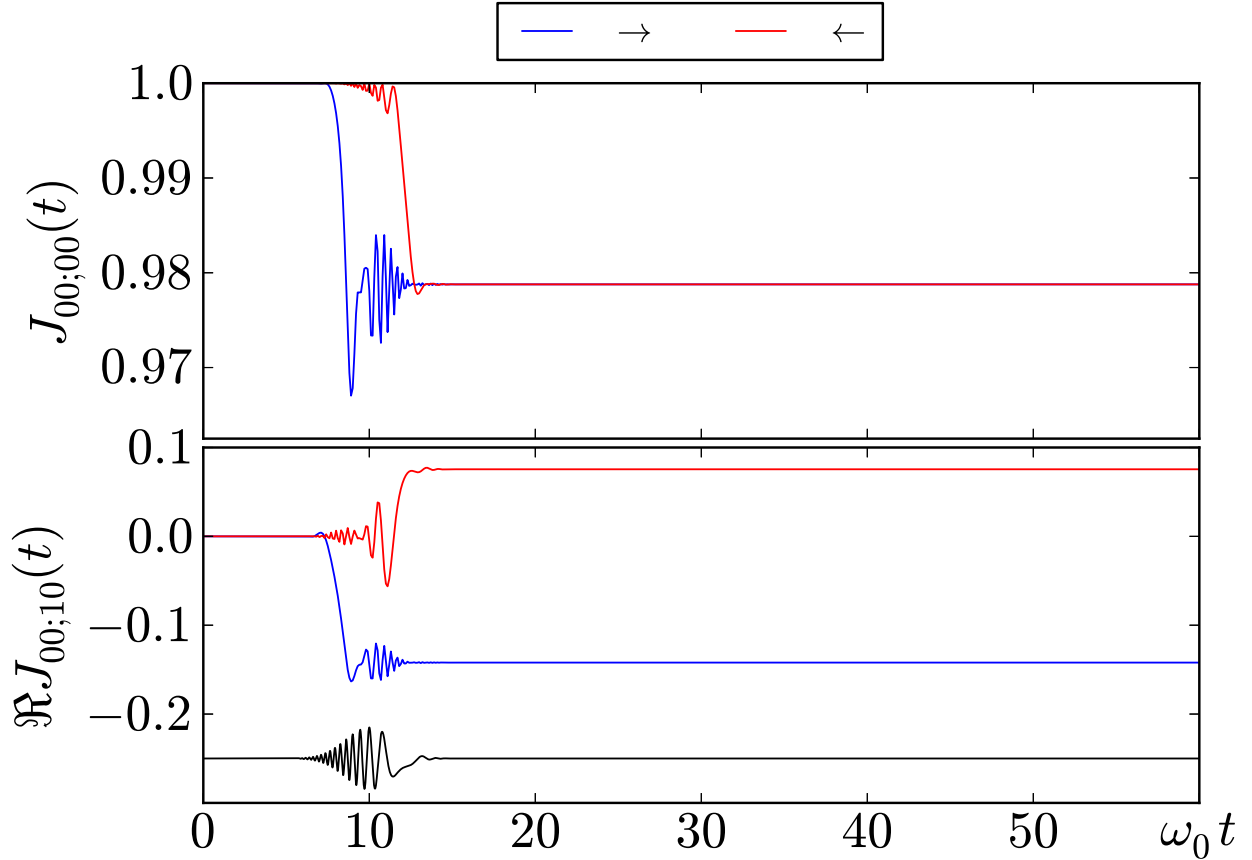}
\end{tabular}
\caption{\textbf{Phase dependence in unitary dynamics.}
$J_{00;00}(t)$ (upper panel) and $J_{00;10}(t)$ (lower panel) induced by the chirped 
pulse in Eq. (\ref{equ:pulse}) with  $\omega_0 t_0=10$, $\Omega_{\mathrm{L}} = 10\omega_0$,
$\Delta t \omega_0 = 4$ and $\sqrt{|S(\omega_0)|}/\mathcal{E}_0 = 0.003415$ with 
$\chi=2.5$ [($\rightarrow$) dark blue curve] and  $\chi=-2.5$ [($\leftarrow$) light red curve].
For reference, we have added at the bottom of the lower panel the laser pulse as a function
of time (black curve).
}
\label{fig:J00nmUnitary}
\end{figure}

 In order to make a connection with the general weak field results derived in
Refs.~\citenum{SAB10}~and~\citenum{PYB13},  we consider
Eqs.~(\ref{Equ:DefPropUnitEvoJnn00})-(\ref{Equ:DefPropUnitEvoJnn11})
in the weak field regime, $\sqrt{|S(\omega_0)|}/\mathcal{E}_0\ll 1$, to get:
\begin{align}
\label{Equ:DiagProp}
J_{00;00}&(\infty)= 1 - J_{00;11}(\infty) ,
\\
J_{00;10}&(\infty) =
\frac{1}{\mathcal{E}_0} \int_{-\infty}^{\infty} \mathrm{d}s \,
\mathrm{e}^{-\mathrm{i}\omega_0 s} E(s),
\\
J_{00;11}&(\infty) =
\frac{1}{\mathcal{E}_0^2}\left|\int_{-\infty}^{\infty} \mathrm{d}s \,
\mathrm{e}^{-\mathrm{i}\omega_0 s} E(s) \right|^2.
\end{align}
These expressions correspond to the typical results in lowest order
perturbation theory in the field amplitude \cite{MB03}.
Clearly, phase dependence is not manifest in the long-time diagonal
terms $J_{00;00}(\infty)$ or $J_{00;11}(\infty)$.
Sample results are shown in Fig.~\ref{fig:J00nmUnitary}, it is clear that no phase dependence is present 
after the pulse is over. Hence, the long time regime is defined as \emph{after} 
the pulse is over.
As discussed below, this is no longer the case when environmental
effects are considered.

Interestingly, for this particular case,  the weak field condition
$\sqrt{|S(\omega_0)|}/\mathcal{E}_0 = \sqrt{|S(\omega_0)|}/\sqrt{2m\hbar \omega_0}\ll 1$
implies that, for the  same value of $\sqrt{|S(\omega_0)|}$, one could
be  in a weak or strong field regime depending on the coupling constant
($k=m \omega_0^2$) and on the mean level spacing $\hbar \omega_0$.

\section{Non-unitary evolution and One-photon Phase-control}
\label{sec:ControlResults}
Consider then the case of dynamics and control in an open system.
We treat the dissipative dynamics using the influence functional approach
\cite{FV63}.
The starting condition is obtained by coupling the central system S to an external
system B which is represented by an infinite number of freedoms \cite{FV63}.
This system B can be, e.g., a thermal bath, the vibrational modes of a molecular complex,
blackbody radiation, etc.
The Hamiltonian of the the system S plus B can be written as
\begin{equation}
\label{equ:TotHamiltonian}
\hat{H} = \hat{H}_{\mathrm{S}} + \hat{H}_{\mathrm{B}} + \hat{H}_{\mathrm{SB}}
+ \hat{V}_{\mathrm{L}}
\end{equation}
where $\hat{H}_{\mathrm{S}} + \hat{V}_{\mathrm{L}}$ are given in Eq.~(\ref{equ:HS}), 
$\hat{H}_{\mathrm{B}}$ is the Hamiltonian of the thermal bath and $\hat{H}_{\mathrm{SB}}$
describes the interaction of the system with B.
We assume that system B is composed of a collection of harmonic oscillators
with masses $m_j$, frequencies $\omega_j$ and coupled linearly to S with
constant couplings $c_j$ \cite{CL83,GSI88,Wei12}.

After tracing over the environment  the influence of the bath
on the time evolution of the system S is described via the spectral density
$J(\omega)$, given by \cite{CL83,GSI88,Wei12}
$
J(\omega) = \pi \sum_{j=1}^{\infty} \frac{c_j^2}{2m_j\omega_j}\delta(\omega-\omega_j).
$ 
Once $J(\omega)$ is fixed, one can express the relaxation process by
means of the dissipative kernel 
$\gamma(s) = \frac{2}{m} \int_0^{\infty} \frac{\mathrm{d}\omega}{\pi}
\frac{J(\omega)}{\omega}\cos(\omega s)$,
while the decoherence process induced by thermal fluctuations can be described
by the decoherence kernel
$
K'(s) =\int_0^{\infty} \frac{\mathrm{d}\omega}{\pi}
J(\omega)\mathrm{coth}\left(\frac{\hbar \beta \omega}{2}\right)\cos(\omega s).
$ 
Here $\beta= 1/k_{\mathrm{B}}T$ and $T$ denotes the temperature of the
thermal bath.

Below, we primarily employ the most commonly used spectral density,
the Ohmic spectral density with a finite Drude cutoff $\omega_{\mathrm{D}}$,
\begin{align}
\label{equ:JwTB}
J_{\mathrm{Ohm}}(\omega) &= m \gamma\omega\,
\omega_{\mathrm{D}}^2/\left(\omega^2 + \omega_{\mathrm{D}}^2\right),
\end{align}
where $\gamma$ is the strength coupling constant to the thermal bath.
This spectral density generates the dissipative kernel
$\gamma(s) = \gamma \omega_{\mathrm{D}} \exp\left(-\omega_{\mathrm{D}} |t|\right)$.
In the limit when the cutoff frequency $ \omega_{\mathrm{D}}$ tends to infinity,
$\gamma(s) \rightarrow 2 \gamma \delta(s)$, corresponding to Ohmic
dissipation.

\subsection{Description of the Initial State}
\label{subsect:IniSta}
Often, the initial state of S + B is assumed to be uncorrelated product \cite{FV63,CL83},
i.e. $\hat{\rho}_{\mathrm{S + B}} = \hat{\rho}_{\mathrm{S}} \otimes \hat{\rho}_{\mathrm{B}}$.
However, this is not a sensible  approximation\cite{GSI88} since it implies that 
system-bath interaction is turned on suddenly when the observation begins.
In the absence of a specific  initial system preparation,
the most likely initial state for S+B is thermal. In this case
if the coupling to the environment is strong, then the equilibrium state is given
by \cite{GSI88}
$
\hat{\rho}_{\mathrm{S},\beta} = Z^{-1}_{\beta} \mathrm{tr}_{\mathrm{B}}
\exp\left[-\left(
\hat{H}_{\mathrm{S}} +
\hat{H}_{\mathrm{B}} +
\hat{H}_{\mathrm{SB}}\right)
\beta\right],
$ 
where $Z_{\beta} $ is the total partition function and tr$_{\mathrm B}$ denotes a trace over the bath.

For our particular case, $ \hat{\rho}_{\mathrm{S},\beta} $ can be expressed
analytically in terms of the effective Hamiltonian \cite{GWT84,HR85,PTB13}
$
\hat{H}_{\mathrm{eff}} = \frac{1}{2 m_{\mathrm{eff}}} \hat{p}^2
+ \frac{1}{2}m_{\mathrm{eff}} \omega_{\mathrm{eff}}^2 \hat{q}^2,
$
with effective mass
$
m_{\mathrm{eff}} =
\omega_{\mathrm{eff}}^{-1} \sqrt{\langle p^2\rangle\langle q^2\rangle^{-1}},
$ and effective frequency $
\omega_{\mathrm{eff}} = 2(\hbar \beta_{\mathrm{TB}})^{-1} \mathrm{arccoth}\left(
\frac{2}{\hbar}\sqrt{\langle p^2\rangle\langle q^2\rangle}\right),
$
being
$\langle q^2\rangle =
\mathrm{tr}_{\mathrm{S}}(\hat{q}^2 \hat{\rho}_{\mathrm{S},\beta})$  and
$\langle p^2\rangle =
\mathrm{tr}_{\mathrm{S}}(\hat{p}^2 \hat{\rho}_{\mathrm{S},\beta})$,
the equilibrium second moments\cite{GWT84,GSI88}.
This $\hat{H}_{\mathrm{eff}}$ definition allows us to express $\hat{\rho}_{\mathrm{S},\beta}$ as
\begin{equation}
\label{equ:EffTherStat}
\hat{\rho}_{\mathrm{S},\beta} = Z_{\beta}^{-1} \sum_{n=0}^{\infty}
\exp(-E_{n_{\beta}} \beta_{\mathrm{TB}}) | n_{\beta} \rangle \langle n_{\beta} |,
\end{equation}
where $\{E_{n_{\beta}} = \hbar \omega_{\mathrm{eff}}(n + \frac{1}{2})\}$ are the
eigenvalues and $\{| n_{\beta} \rangle\}$ the eigenstates of the effective Hamiltonian
$\hat{H}_{\mathrm{eff}}$.
At high temperature, $\hbar \omega_0 \beta \ll 1$ and
$\frac{1}{2} \hbar \gamma\beta \ll 1$,
$m_{\mathrm{eff}}$ and $\omega_{\mathrm{eff}}$ approach their bare values
$m$ and $\omega_0$, respectively and, therefore, $\hat{\rho}_{\mathrm{S},\beta}$
approaches the canonical distribution\cite{GWT84,HI05,PTB13}.
By contrast, at low temperatures $\hbar \omega_0 \beta \gg 1$ and
$\frac{1}{2} \hbar \gamma \beta \gg 1$,
$m_{\mathrm{eff}}$ and $\omega_{\mathrm{eff}}$ deviate from the bare 
values $m$ and $\omega_0$ due to damping\cite{GWT84,HI05,PTB13}.
This deviation from the canonical distribution implies that the initial equilibrium
state [Eq. (\ref{equ:EffTherStat})] contains stationary off-diagonal elements when
it is projected on the eigenbasis of $\hat{H}_{\mathrm{S}}$. Note further that
these off-diagonal elements are expected to vanish in classical mechanics\cite{CTH09}.
Hence, the existence of these terms is a purely quantum mechanical.

\begin{figure}[h]
\includegraphics[width = \columnwidth]{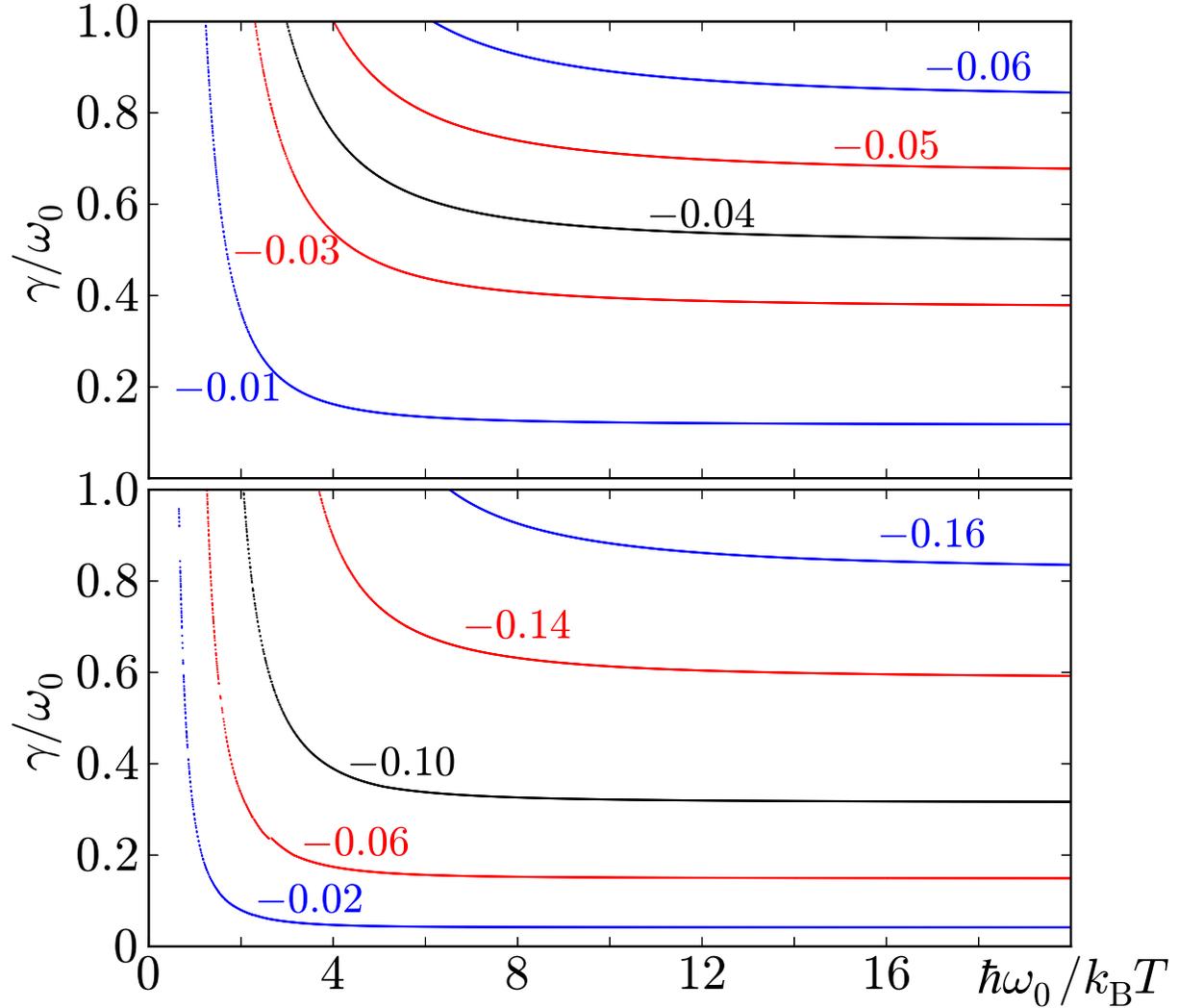}
\caption{\textbf{Stationary coherence generated by the Ohmic spectral density.}
Contours of constant matrix element $\langle 0 | \hat{\rho}_{\mathrm{S},\beta} | 2\rangle$
as a function of $\gamma$ and $T$ for fixed $\omega_{\mathrm{D}}=\omega_0$ (upper panel) and
for $\omega_{\mathrm{D}}=100\omega_0$ (lower panel).}
\label{fig:rho02Equil}
\end{figure}

In order to explore the strength of the off-diagonal elements introduced in Eq.~(\ref{equ:EffTherStat}),
we show, in Fig. \ref{fig:rho02Equil}, the matrix element
$\langle 0 | \hat{\rho}_{\mathrm{S},\beta} | 2\rangle$ as a function of the coupling
constant $\gamma$ and temperature $T$ for the non-Markovian regime,
$\omega_{\mathrm{D}}=\omega_0$ (upper panel) and for the Markovian regime,
$\omega_{\mathrm{D}}=100\omega_0$ (lower panel).
In both cases, for fixed $\gamma$ 
the matrix element $\langle 0 | \hat{\rho}_{\mathrm{S},\beta} | 2\rangle$
is larger at low temperatures, whereas for fixed temperature it is larger for larger 
coupling constant $\gamma$.
From Fig.~(\ref{fig:rho02Equil}) one can also see that the larger the cutoff $\omega_{\mathrm{D}}$,
the larger the matrix element $\langle 0 | \hat{\rho}_{\mathrm{S},\beta} | 2\rangle$ at fixed
$\gamma$ and $T$.
This is consistent with the fact that the more Markovian the system,
the larger is the
decay rate, and therefore the larger the \emph{effective} coupling to the bath (for an 
extended discussion on this parameter dependence see Ref.~\citenum{PTB13}).

Note that increasing the value of 
$\langle 0 | \hat{\rho}_{\mathrm{S},\beta} | 2\rangle$
would imply a stronger contribution of these off-diagonal terms to the subsequent
dynamics. 
However,  in general, it would also imply fast relaxation, and therefore very fast loss 
of phase information.
That is, coupling of the system to the bath plays two roles in phase control,
one to enhance the phase control and one to cause relaxation. 
Below, in order to explicitly explore the system time-evolution and elucidate the 
contributions to phase control, we choose a moderate coupling constant, 
$\gamma= 0.1\omega_0$ and low temperature $\hbar\omega_0/k_{\mathrm{B}}T=40$.

\subsection{Effect of the Initial Correlations}
Based on the discussion for the unitary case, one can identify the deviations from the canonical
distribution, and the related presence of off-diagonal
elements in the initial density matrix, as the first contribution of the bath to one photon
phase control.
This implies that when laser excitation takes place, it finds \emph{off-diagonal} elements
of $\hat{\rho}_{\mathrm{S}}$, where laser phase information could be encoded (see below).
This contribution occurs then on the time scale of the laser field\cite{PYB13}.

Before considering the time evolution of the off-diagonal elements, it is useful to consider
which of these terms are non-zero.
From the symmetry of the resulting equilibrium state [see Eq. (\ref{equ:EffTherStat})],
we expect the non-vanishing off-diagonal elements 
$\langle n | \hat{\rho}_{\mathrm{S},\beta} (0)| m \rangle$ that satisfy: 
$\mathrm{parity}(\langle n |q\rangle) \times \mathrm{parity}(\langle q | m \rangle) = \mathrm{even}$;
this is the case in, e.g., $\langle 0 | \hat{\rho}_{\mathrm{S},\beta} (0)| 2 \rangle$ or 
$\langle 1 | \hat{\rho}_{\mathrm{S},\beta} (0)| 3 \rangle$.
Hence, the bath is not able to induce stationary off-diagonal elements such as 
$\langle 0 | \hat{\rho}_{\mathrm{S},\beta} (0)| 1 \rangle$ or 
$\langle 1 | \hat{\rho}_{\mathrm{S},\beta} (0)| 2 \rangle$.
Although a formal analysis of this fact could be carried out on the basis of the transitions
induced by the bath after being traced out, for our proposes it suffices to consider
this result as a consequence of the symmetries of the equilibrium state.
Note that a similar analysis needs to be carried out for each particular system. 

To demonstrate the phase dependence due to the stationary coherences in our model system, consider
the propagating function element responsible 
for the $\langle 0 | \hat{\rho}_{\mathrm{S}} (0)| 2 \rangle =
\langle 0 | \hat{\rho}_{\mathrm{S},\beta}| 2 \rangle$ to the time-evolution of the ground 
state
$\langle 0 | \hat{\rho}_{\mathrm{S}} (t)| 0 \rangle$,
\begin{equation}
\label{equ:J0002}
\begin{split}
\hspace{-0.15cm}J_{00;20}&(t) =
 \sqrt{\frac{8\pi^2}{q_0^8\hbar^4\det\mathsf{M}}}
 \frac{1}{N(t)}
 \mathrm{e}^{\frac{1}{2} \mathsf{E}^{\mathrm{T}} \mathsf{A} \mathsf{E}}
  \left\{(\mathsf{A}_{33} - q_0^2)\hbar^2 \right.
\\
&\left. - 2\left[(\mathsf{A}_{13} -\mathsf{A}_{23} )E_-(t)
(\mathsf{A}_{33} -\mathsf{A}_{34} )\right]^2 E_+(t) \right\},
\end{split}
\end{equation}
where $q_0 = \sqrt{\hbar/m\omega_0}$,
$N(t) = 2\pi \hbar \frac{1}{m}|G_+(t)|\sqrt{2\pi \langle \hat{q}^2\rangle}$,
 and
$\mathsf{E} = \left( E_+(t),  E_-(t) \right)$ with
$
E_+(t) = \int_0^t  \mathrm{d}s E(s) G_+(t-s)/G_+(t),
$ $
E_-(t) = \int_0^t \mathrm{d}s E(s) G_-(s)/G_-(t).
$ 
Here $G_{+}(s)$ denotes the inverse Laplace transform of
$G(z) = \left(z^2 + z\gamma(z) + \omega_0^2 \right)^{-1}$,
being $\gamma(z)$ the Laplace transform of the dissipative kernel
$\gamma(s)$ \cite{GSI88}.
$G_{-}(s)$ is related to $G_{+}(s)$ by means of the relation
$\frac{G_{-}(s)}{G_{-}(t)} = \dot{G}_+(t-s)
- \frac{G_+(t-s)}{G_+(t)} \dot{G}_+(s)$.
As required, in the limit of vanishing coupling to the bath $\gamma \rightarrow 0$,
$G_{+}(s) = G_{-}(s) =
\frac{1}{2\mathrm{i} \omega_{0}}\left[ \exp(\lambda_1 t) - \exp(\lambda_2 t)\right]$
with $\lambda_{1,2} = \pm \mathrm{i} \omega_0$
and
$
\mathsf{E}^{\mathrm{T}} \mathsf{A} \mathsf{E} = -2|\mathcal{E}(t)|^2/\mathcal{E}_0^2,
$
the expressions in 
Eqs.~(\ref{Equ:DefPropUnitEvoJnn00}-\ref{Equ:DefPropUnitEvoJnn11}) are recovered.
In Appendix~\ref{app:EvoDenOpe}, we present the explicit expressions for 
$\mathsf{A}$ and $\mathsf{M}$\footnote{A Mathematica 8.0 script with the numerical 
implementation of the results for the Ohmic spectral density can be found at 
\href{http://gfam.udea.edu.co/~lpachon/scripts/}
{http://gfam.udea.edu.co/$\sim$lpachon/scripts/oqsystems}.}.

In the long time regime,
\begin{equation}
\label{equ:DefPropNonUnitEvoJ0020}
\begin{split}
J_{00;20}(t) &\sim \frac{\hbar^3(\langle p^2 \rangle - m \omega_0 \hbar) A^2(t) E^2_-(t)}{
m^2 \langle q^2 \rangle \dot{A}^2(t)\sqrt{2m \omega_0(4m \omega_0 \langle q^2 \rangle -\hbar)}}
\\
&\times \frac{\exp\left(\frac{1}{2} \mathsf{E}^{\mathrm{T}} \mathsf{A} \mathsf{E}\right)}
{\sqrt{4m^4 \langle q^2 \rangle\frac{\dot{A}^2(t)}{G^2_+(t)}
+ \hbar^2(\langle p^2 \rangle - m \omega_0 \hbar)}},
\end{split}
\end{equation}
with
\begin{align}
\label{equ:LongTimeReg}
\mathsf{E}^{\mathrm{T}} \mathsf{A} \mathsf{E}
&\sim -\frac{2E_-^2(t)}{m \omega_0 \hbar - \langle p^2 \rangle
-m^2\frac{\dot{A}^2(t)}{A^2(t)}\langle q^2 \rangle},
\end{align}
where $A(t)$ is the antisymmetric correlation function
$A(t) = \frac{1}{2\iu}\langle \hat{q}(t) \hat{q} - \hat{q}\hat{q}(t)\rangle 
= -\frac{\hbar}{2m} G_+(t)$ 
(see Appendix~\ref{app:EvoDenOpe}).
In contrast to the result in Eq.~(\ref{equ:GenFouTraField}),  
Eq.~(\ref{equ:LongTimeReg}) is 
independent of $E_+(t)$; this is a result of the time-symmetry breaking
anticipated above.
Hence, it is clear that in this case, $\mathsf{E}^{\mathrm{T}} \mathsf{A} \mathsf{E}$ 
differs from the squared modulus of the Fourier transform and therefore phase
information can be stored, \emph{even} in the diagonal elements 
of the propagating function.
\begin{figure}[h]
\includegraphics[width = \columnwidth]{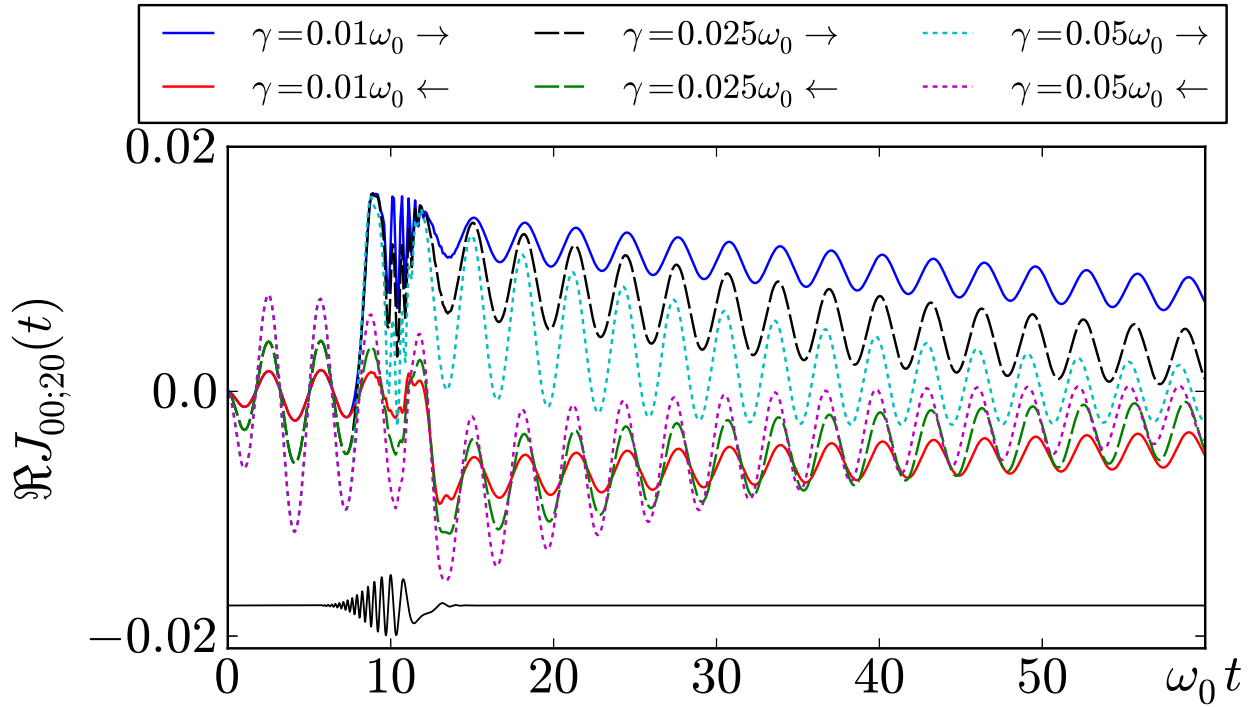}
\caption{\textbf{
Time evolution of the propagating function elements associated with stationary coherences.} 
$J_{00;20}(t)$ induced by the chirped pulse in Eq.~(\ref{equ:pulse})
with $\omega_{\mathrm{D}} = \omega_0$ and $\hbar \omega_0/k_{\mathrm{B}}T = 40$.
Here $\omega_0 t_0=10$, $\Omega_{\mathrm{L}} = 10\omega_0$,
$\Delta t \omega_0 = 4$ and $\sqrt{|S(\omega_0)|}/\mathcal{E}_0 = 0.003415$ with 
$\chi=2.5$ [($\rightarrow$) continuous blue, 
dashed black and dotted cyan curves] and  $\chi=-2.5$ [($\leftarrow$) continuous red, 
dashed green and dotted magenta curves]
for various values of the coupling constant $\gamma$.
}
\label{fig:J0020}
\end{figure}

This being the case, the spectral phase information contained in the propagating functions element $J_{00;20}$
can now affect the dynamics of the expectation value in Eq.~(\ref{ExpValuObsProFun}).
Figure~\ref{fig:J0020} shows $J_{00;20}$ as a function of time for $\chi=2.5$
(continuous blue, dashed black and dotted cyan curves) and 
$\chi=-2.5$ (continuous red, dashed green and dotted magenta curves) and for three different 
values of the coupling constant $\gamma=0.01\omega_0$, $\gamma=0.025\omega_0$ and
$\gamma=0.05\omega_0$.

Remarkably, in Fig.~\ref{fig:J0020}, $J_{00;20}(t)$ shows a time dependence \textit{before} the 
laser excitation occurs, resulting from the time dependence of the antisymmetric 
correlation function $A(t)$.
This is a manifestation of an \emph{an incoherent flux at equilibrium} between eigenstates.
The detailed origin of this incoherent flux is well beyond the scope of this work and will be 
discussed elsewhere. 

\subsection{Effect of Time-reversal-symmetry Breaking}
The discussion above focused upon stationary features of non-unitary evolution that assist OPPC.
Here we consider dynamical features (time-reversal-symmetry 
breaking) in non-unitary evolution that enhance OPPC.
For this propose it suffices to study one of the propagating function elements.
For example, the $J_{00;00}(t)$ element reads
\begin{equation}
\label{equ:DefPropNonUnitEvoJ0000}
\begin{split}
J_{00;00}(t) =
 \sqrt{\frac{1}{ q_0^4}\frac{2^4}{\det\mathsf{M}}}
 \frac{1}{N(t)}
\exp\left(\frac{1}{2} \mathsf{E}^{\mathrm{T}} \mathsf{A} \mathsf{E}\right),
\end{split}
\end{equation}
Based on Eq.~(\ref{equ:LongTimeReg}), it is clear that
the propagating element $J_{00;00}(t)$ depends on the electric field phase
in the long time regime for the open system case,
thus allowing for the encoding of phase information.

\begin{figure}[h]
\includegraphics[width = \columnwidth]{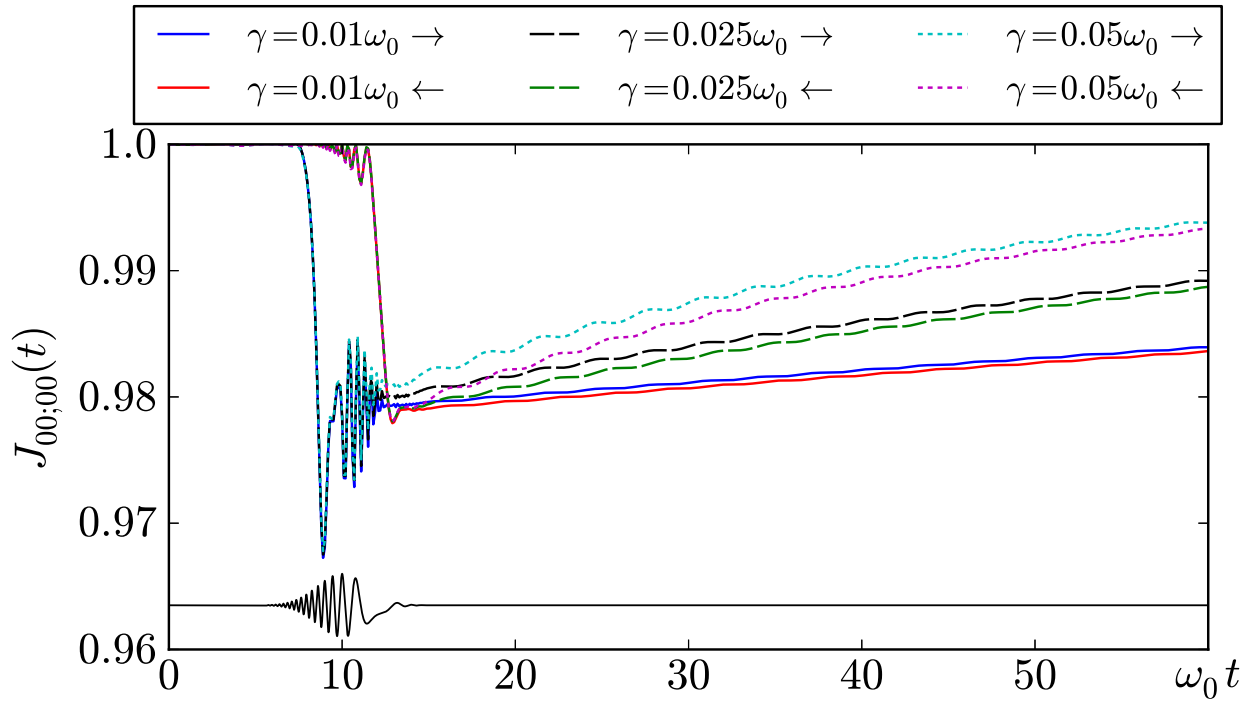}
\caption{\textbf{
Time evolution of the propagating function elements associated to populations.} 
$J_{00;00}(t)$ induced by the chirped pulse in Eq.~(\ref{equ:pulse}).
Parameter values as in Fig.~\ref{fig:J0020}.}
\label{fig:Jnn00}
\end{figure}

Fig.~\ref{fig:Jnn00} shows $J_{00;00}(t)$ using the
same parameters as in Fig.~\ref{fig:J0020}.
For the unitary case [see Fig.~\ref{fig:J00nmUnitary}], after the pulse is over there is
no phase information in $J_{00;00}(t)$.
However, for the non-unitary dynamics the effect of the phase 
persists in the long time regime, albeit diminished by the underlying 
incoherent thermal process. 
Further, the magnitude of the phase effect in $J_{00;00}(t)$, for short times after the 
pulse is over, is larger for the larger values of the coupling to the environment.
However, in the long time regime, due to the equilibration, the
phase information is lost.

\subsection{Control of Observables}
Consider then, as an example, control over the population of the
$n$-th system eigenstate,  e.g.,
the population of the ground state, with $\hat{O}  = |0\rangle \langle 0|$.
Under unitary dynamics, and for an incoherent initial state, the time evolution of the
ground state population is completely generated by the propagating function element 
$J_{00,00}(t)$ (see Fig.~\ref{fig:J00nmUnitary}), and, as described above, no 
phase control is possible.

\begin{figure}[h]
\includegraphics[width = \columnwidth]{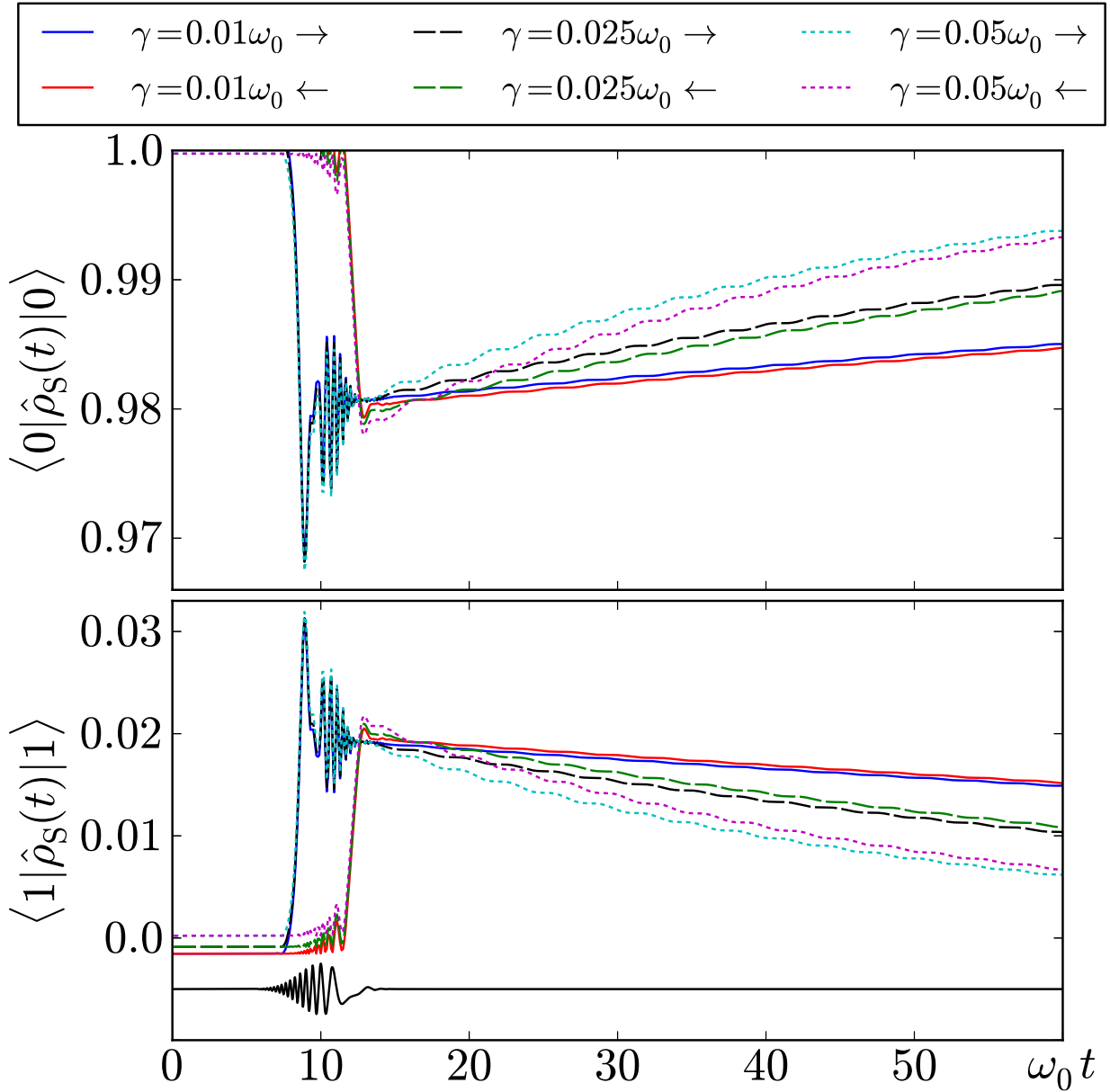}
\caption{
\textbf{Influence of the coupling constant for the Ohmic spectral density.}
Time evolution of $\langle 0|\hat{\rho}(t)|0\rangle$ (upper panel) and 
$\langle 1|\hat{\rho}(t)|1\rangle$ (lower panel) induced by the chirped pulse in 
Eq.~(\ref{equ:pulse}).
Parameter values as in Fig.~\ref{fig:J0020}.}
\label{fig:gDrho00}
\end{figure}

The situation is different in the presence of an environment.
The upper panel of Fig.~\ref{fig:gDrho00} shows the time evolution of the 
ground state population $\langle 0|\hat{\rho}_{\mathrm{S}}(t)|0\rangle$ using the same parameters
as in Fig.~\ref{fig:J0020}.
Here phase dependence of
a quantity that commutes with the bare Hamiltonian $\hat{H}_{\mathrm{S}}$, is evident.
Note that the full pulse effectively spans a time of $10 \omega_0^{-1}$.
Hence, although phase control must be lost over long times due to the environmental
coupling, the control still survives over an extended time after 
the pulse is over, i.e., in a ``non-equilibrated regime".

The lower panel of Fig.~\ref{fig:gDrho00} shows the time-evolution dynamics
of the first excited state $\langle 1|\hat{\rho}(t)|1\rangle$.
The characteristics of the phase dependence are similar to those for the
ground state dynamics, the smaller phase amplitude and slower decay being
the only noticeable differences.
This arises from the very small initial population in 
$\langle 1|\hat{\rho}_{\mathrm{S}}(t)|1\rangle$ 
(0.000446 for $\gamma=0.01\omega_0$,
0.00113 for $\gamma=0.025\omega_0$ and 
0.00223 for $\gamma=0.05\omega_0$)
and the fact that for the times where the amplitude
of the phase dependence is moderate, immediately after the pulse, the propagating
elements $J_{11;\nu\mu}$ are small. 
In particular, the term responsible for the 
transfer of population from the ground state, $J_{11;00}$, behaves as 
$\sim \frac{1}{2} \mathsf{E}^{\mathrm{T}} \mathsf{A} \mathsf{E}
\exp\left(\frac{1}{2} \mathsf{E}^{\mathrm{T}} \mathsf{A} \mathsf{E}\right)$ whereas
$J_{00;00}$ behaves as  $\sim 
\exp\left(\frac{1}{2} \mathsf{E}^{\mathrm{T}} \mathsf{A} \mathsf{E}\right)$
[see Eq.~(\ref{equ:DefPropNonUnitEvoJ0000})].

In both panels of Fig.~\ref{fig:gDrho00} one observes that, after the pulse is over, the state populations
reach certain chirp-dependent values and then relax.
Comparing the results in Fig.~\ref{fig:gDrho00} with the time evolution of the $J_{00;00}$ element in 
Fig.~\ref{fig:Jnn00}, shows that the phase dependence, for this set of parameters,
is primarily dictated by  $J_{00;00}$, with no appreciable influence from the 
stationary coherence $\langle 0|\hat{\rho}_{\mathrm{S}}(0)|2\rangle$.
This is a result of the small value of $J_{00;20}(t)\langle 0|\hat{\rho}_{\mathrm{S},\beta}|2\rangle$, 
which is the leading off-diagonal-contribution to the dynamics of $\langle 0|\hat{\rho}_{\mathrm{S}}(0)|0\rangle$.
In particular, for 
$\gamma = 0.01\omega_0$ we have that 
$\langle 0|\hat{\rho}_{\mathrm{S},\beta}|2\rangle = -0.00088$;
while for  
$\gamma= 0.025 \omega_0$ and $\gamma = 0.05 \omega_0$,
we have that
$\langle 0|\hat{\rho}_{\mathrm{S},\beta}|2\rangle =  -0.00219$ and 
$\langle 0|\hat{\rho}_{\mathrm{S},\beta}|2\rangle =  -0.00435$, respectively. The off-diagonal 
element contribution may well be larger for other kinds of systems and couplings.

Given our goal of significant phase control, Fig.~\ref{fig:rho02Equil} suggests that we increase, 
e.g., the ratio $\gamma/\omega_0$.
However, doing so would also induce faster decay rates, and the phase dependence would 
quickly vanish.
Alternately, since the thermal energy at room temperature is $\sim 1/40$~eV and
energy of the optical transitions are between 1.6 to 3.4~eV, increasing the ratio 
$\hbar \omega_0/k_{\mathrm{B}}T$ could be a more promising 
alternative to enhance control.
However, as seen from Fig.~\ref{fig:rho02Equil}, for 
$\hbar \omega_0/k_{\mathrm{B}}T > 10$, $\langle 0|\hat{\rho}_{\beta}|2\rangle$ 
becomes basically temperature-independent.
Thus environmentally assisted phase control has the fundamental challenge 
that the terms that assist control also induce, simultaneously, equilibration with the bath.

\subsection{Influence of the Non-Markovian Character of the Dynamics}
\begin{figure}[h]
\begin{tabular}{cc}
\includegraphics[width = \columnwidth]{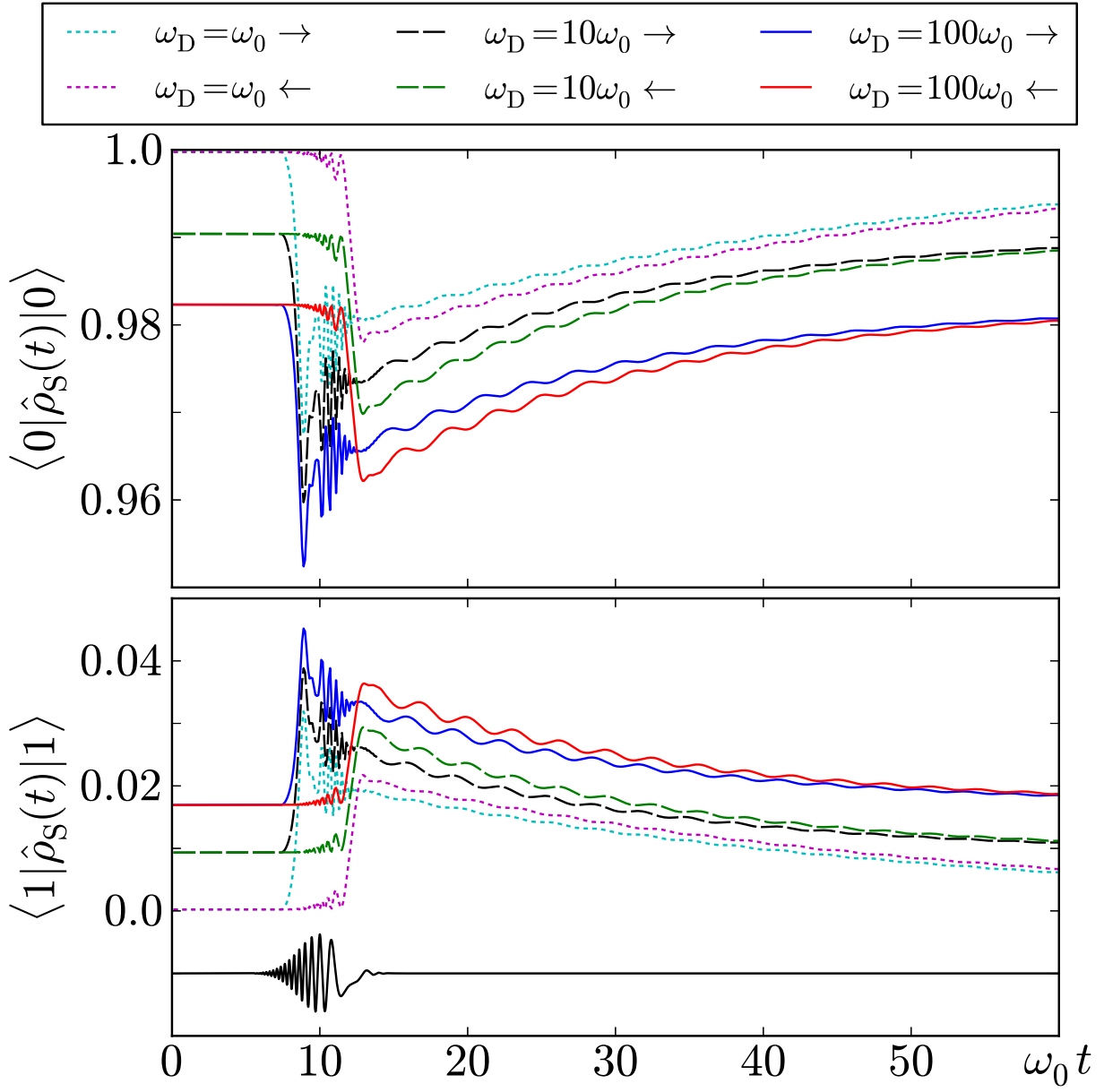}
\end{tabular}
\caption{\textbf{Influence of the non-Markovian character of the dynamics.}
Time evolution of $\langle 0|\hat{\rho}_{\mathrm{S}}(t)|0\rangle$ (upper panel) and 
$\langle 1|\hat{\rho}_{\mathrm{S}}(t)|1\rangle$ (lower panel) with 
$\gamma = 0.05\omega_0$ and $\hbar \omega_0/k_{\mathrm{B}}T = 40$
for different values of the cutoff frequency $\omega_{\mathrm{D}}$.
The value of the laser parameters are as in Fig.~\ref{fig:J0020}.
}
\label{fig:wDrho00}
\end{figure}

In general, open quantum systems undergo non-Markovian provide that the spectrum
of the thermal fluctuations is not flat (coloured noise\cite{HJ07}).
Formally, the Markovian regime is reached when the spectral density has a linear monotonic 
behaviour and thermal energy $k_{\mathrm{B}}T$ is the larger energy scale of the system.
For the Ohmic spectral density in Eq.~(\ref{equ:JwTB}), increasing $\omega_{\mathrm{D}}$
is known to bring the system closer to the Markovian limit.
To examine the non-Markovian  effects we show, in Fig.~\ref{fig:wDrho00}, the time evolution 
of the ground state and first excited state, generated by the laser pulse for different values
of the cutoff frequency $\omega_{\mathrm{D}}$.
The amplitude of the phase effect, for short times after the pulse is 
over, is seen to be larger for the Markovian cases ($\omega_{\mathrm{D}}=10\omega_0$ 
and $\omega_{\mathrm{D}}=100\omega_0$) than for the non-Markovian case
($\omega_{\mathrm{D}}=\omega_0$).
This reflects the stronger effective system-bath coupling in the Markovian case
\cite{PTB13}.
As a consequence, the time reversibility of the unitary dynamics is lost more effectively 
in the non-Markovian case.
However, in the long time regime the non-Markovian dynamics, due to its weaker 
effective coupling to the bath, has a larger phase-effect, albeit the overall phase control is
very small.

\subsection{Effect of the a Different Spectral Density}
Although the main lines of the above discussion are general and intuitive, the numerical results 
above pertain to the Ohmic spectral density [\ref{equ:JwTB}].
Motivated by our recent observation\cite{PB13c} that the effect of the vibrations and the solvent--
in systems such as dye molecules, amino acid proteins and some photochemical 
systems (e.g., rhodopsin and green fluorescence proteins)-- could be effectively described
by a sub-Ohmic spectral density; we consider this alternate spectral density below.
Note that sub-Ohmic spectral densities are characterized by slow decay rates, even 
in the strong coupling regime\cite{PB13c}.
Hence, this spectral density may well be of some interest in terms of phase-control, 
where it may result in a prolonged phase-control effect. 

The sub-Ohmic spectral density with exponential decay is given by
\begin{align}
\label{equ:JwTBSO}
J_{\mathrm{sOhm}}(\omega) &= m \gamma \omega_{\mathrm{ph}}^{1-s} \omega^{s} \,
\exp\left( -\omega/\omega_{\mathrm{D}}\right),
\end{align}
with $0<s<1$, and $\omega_{\mathrm{ph}}$ is an auxiliary phononic scale frequency.
For this case the relevant coupling constant is $\gamma \omega_{\mathrm{ph}}^{1-s}$ and
the spectral density generates the dissipative kernel
$\gamma(s) = \frac{4}{\pi} \left(1 + t^2 \omega_{\mathrm{D}}^2 \right)^{-s/2}
\omega_{\mathrm{ph}}^{1-s} \Gamma(s)\cos\left[s \arctan(\omega_{\mathrm{D}} t) \right]$.
In the limit where $\omega_{\mathrm{D}} \rightarrow \infty$,
$\gamma(t) \rightarrow \frac{4}{\pi} t^{-s} \omega_{\mathrm{ph}}^{1-s} \Gamma(s)
\cos\left(\frac{\pi}{2}s \right)$ which, by contrast to the Ohmic case, also leads
to non-Markovian effects.
In the short time regime, $\omega_{\mathrm{D}} t \ll 1$, 
$\gamma(t)/\gamma(0) \sim 1 -\frac{1}{2}s(1+s)(\omega_{\mathrm{D}} t)^2$,
which resembles the functional form of  a Gaussian decay at short times.
In the long time regime, $\omega_{\mathrm{D}} t\gg 1$, 
$\gamma(t)/\gamma(0) \sim \cos(\frac{\pi}{2} s) (\omega_{\mathrm{D}} t)^{-s}$, 
so that the long time decay is only algebraic, $1/t^s$. 

\begin{figure}[h]
\includegraphics[width = \columnwidth]{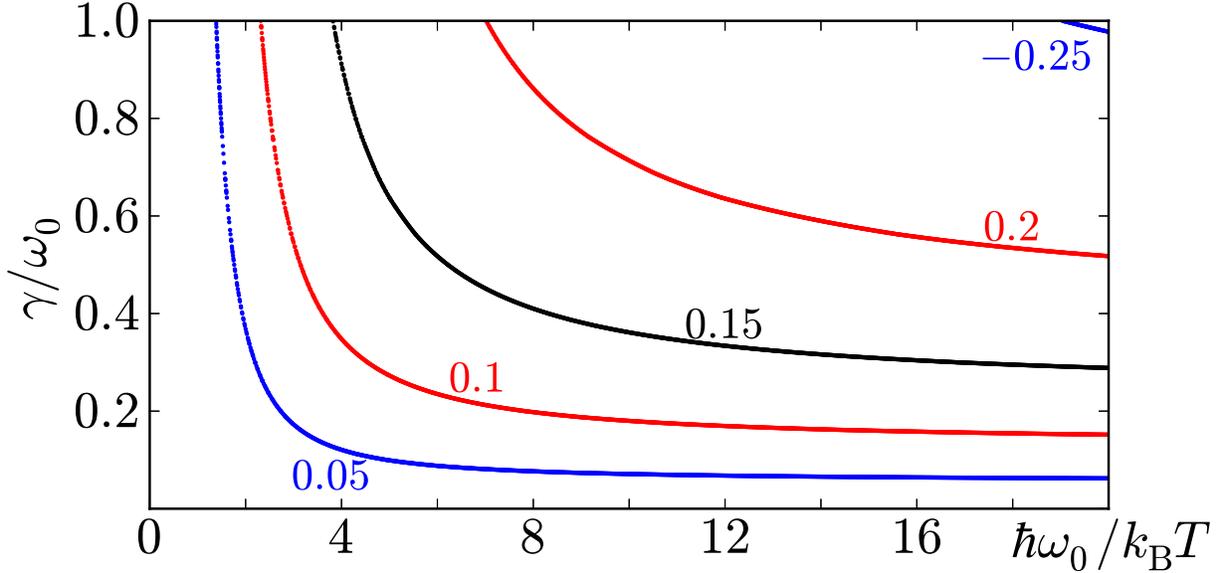}\\
\caption{\textbf{Stationary coherence generated by the sub-Ohmic spectral density.}
Contours of constant matrix element $\langle 0 | \hat{\rho}_{\mathrm{S},\beta} | 2\rangle$
as a function of $\gamma$ and $T$ for fixed $\omega_{\mathrm{ph}}=\omega_0$, 
$\omega_{\mathrm{D}}=\omega_0$ and $s=0.1$.}
\label{fig:rho02EquilSO}
\end{figure}

\begin{figure}[h]
\includegraphics[width = \columnwidth]{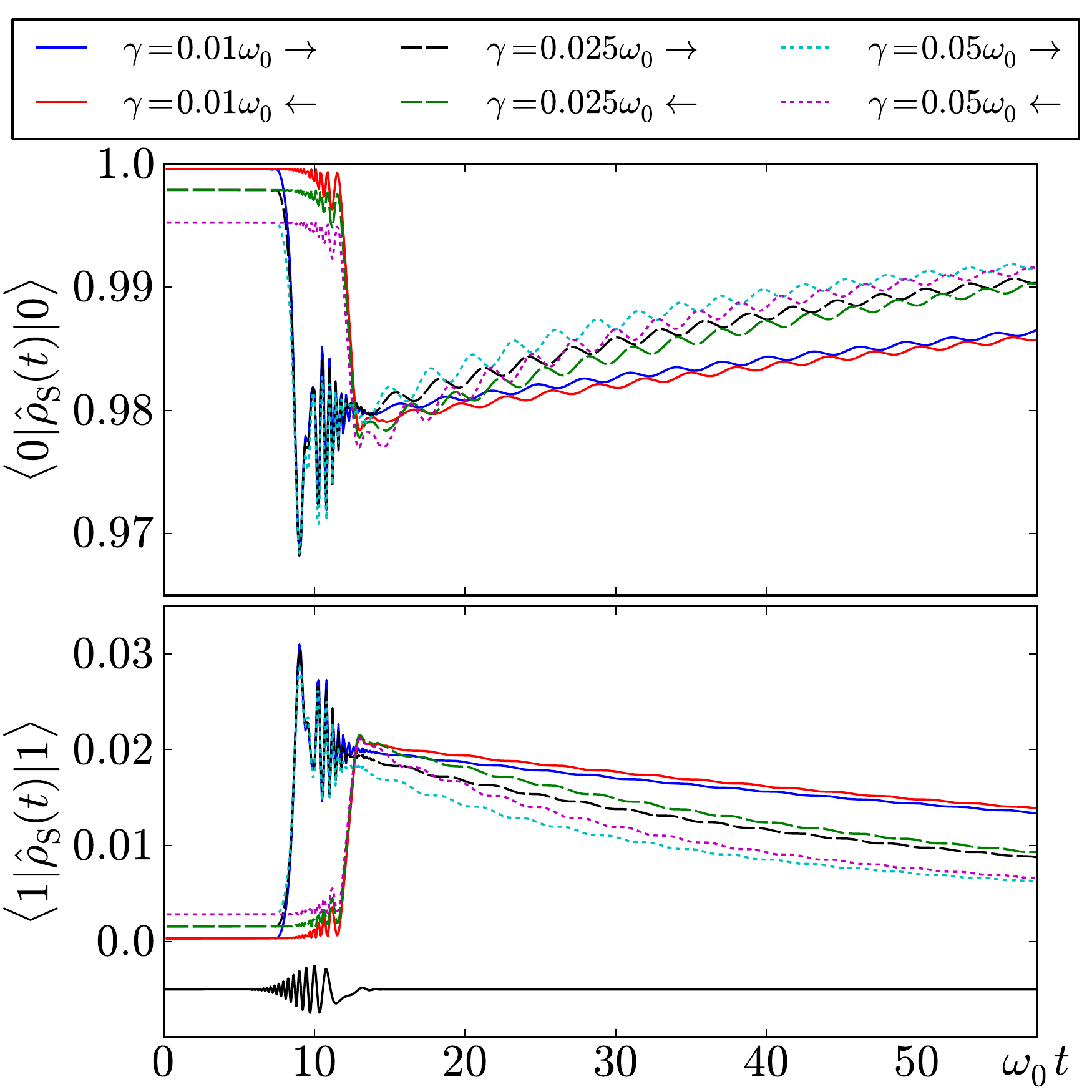}
\caption{
\textbf{Influence of the coupling constant for the sub-Ohmic spectral density.}
Time evolution of $\langle 0 | \hat{\rho}_{\mathrm{S}} | 0\rangle$ induced by the chirped 
pulse in Eq.~(\ref{equ:pulse}).
with and $\hbar \omega_0/k_{\mathrm{B}}T = 40$, $\omega_{\mathrm{ph}}=\omega_0$, 
$\omega_{\mathrm{D}}=\omega_0$ and $s=0.1$.
The value of the laser parameters are as in Fig.~\ref{fig:J0020}.}
\label{fig:Jnn00SO}
\end{figure}

Fig.~\ref{fig:rho02EquilSO} shows the sub-Ohmic analog of the Ohmic calculation
in the upper panel in Fig.~\ref{fig:rho02Equil}, where values of the stationary off-diagonal element  
$\langle 0 | \hat{\rho}_{\mathrm{S},\beta} | 2\rangle$, larger than in the Ohmic case, are seen.
Figure~\ref{fig:Jnn00SO} shows 
the time evolution of the ground and first
excited states, where the system-bath interaction is described by a 
sub-Ohmic spectral density [\ref{equ:JwTBSO}] and where the parameters
are the same as those in Fig.~\ref{fig:gDrho00}
The new feature seen here is the persistence of population oscillations in the ground state
which is also present, although to lesser extent, in the first excited state.
This observation is consistent with the recent finding that sub-Ohmic spectral densities are able
to maintain coherent oscillations for longer times in the spin-boson model\cite{KA13}.

One further feature occurs at short times where, the phase amplitude is 
slightly larger in the sub-Ohmic case than in the Ohmic case.
However, due to the Gaussian-like decay discussed above, this larger phase 
dependence decays faster here than in the Ohmic case (where for short times, 
the decay is exponential).
However, this fast relaxation is compensated by the long-time algebraic decay
(see above), so that some phase-information persists in the system
at long times.

\section{Discussion}
In the  model system examined here, phase controllability was successfully demonstrated, 
primarily mediated by the  
breaking of time-reversal symmetry, which is a completely incoherent process. 
Although we could not observe any significant effect of the off-diagonal elements (stationary
coherences) during the control phase, the fact that it can reach large values 
(see Fig.~\ref{fig:rho02EquilSO}) implies that these terms must be included in any treatment
of phase control and may well be influential in other model systems.

We note that there is a fundamental difference in the nature of control mechanisms arising
via the time-reversal symmetry and the stationary coherences.  
Specifically, the first is an effect that can arise in either classical or quantum mechanics. 
Indeed, proving that the origin of such experimentally observed control is quantum, e.g., 
via a Bell inequality\cite{Sch13}, is far from a trivial task.
By contrast, the off-diagonal elements vanish in classical mechanics\cite{CTH09}. 
Hence, they are quantum in nature, arising from the entanglement between the system
and the bath.
Although in our case the contribution of the off-diagonal elements is of second order in the 
effective coupling to the bath\cite{PTB13}, and therefore is expected to be small, any successful
observation of the role of the off-diagonal elements in OPPC will signal a pure quantum effect.

Indeed, it should be noted that, to date,  there is no experimental demonstration of environmentally
assisted one photon phase control. 
The original\cite{PM&06,Jof07,PM&07} or related experiments\cite{PHM11} in which this mechanism was invoked,
as well as the related computational work\cite{AB13}, showed 
phase control of \textit{cis-trans} isomerization. However, the 
property ``cis-or-trans" consists of a projection onto a spatial domain, and hence is
an operator that does not commute with the system Hamiltonian. As such, one knows\cite{SAB10,PYB13}
that phase control is possible even in the absence of an environment. Rather, the
environment in case of isomerization serves to relax the system into one of the two
isomers. 
The insights afforded by the analysis in this paper should serve to motivate
new studies to experimentally demonstrate one photon phase control.

\acknowledgements
This work was supported by NSERC and by the US Air Force Office of Scientific Research
under contract number FA9550-10-1-0260, and by \emph{Comit\'e para el Desarrollo de la
Investigaci\'on} --CODI-- of Universidad de Antioquia, Colombia under contract number E01651
and under the \textit{Estrategia de Sostenibilidad 2013-2014} and by the \textit{Colombian Institute 
for the Science and Technology Development} --COLCIENCIAS-- under grant number 111556934912.

\bibliography{oppcjcpv2}

\begin{thebibliography}{43}%
\makeatletter
\providecommand \@ifxundefined [1]{%
 \@ifx{#1\undefined}
}%
\providecommand \@ifnum [1]{%
 \ifnum #1\expandafter \@firstoftwo
 \else \expandafter \@secondoftwo
 \fi
}%
\providecommand \@ifx [1]{%
 \ifx #1\expandafter \@firstoftwo
 \else \expandafter \@secondoftwo
 \fi
}%
\providecommand \natexlab [1]{#1}%
\providecommand \enquote  [1]{``#1''}%
\providecommand \bibnamefont  [1]{#1}%
\providecommand \bibfnamefont [1]{#1}%
\providecommand \citenamefont [1]{#1}%
\providecommand \href@noop [0]{\@secondoftwo}%
\providecommand \href [0]{\begingroup \@sanitize@url \@href}%
\providecommand \@href[1]{\@@startlink{#1}\@@href}%
\providecommand \@@href[1]{\endgroup#1\@@endlink}%
\providecommand \@sanitize@url [0]{\catcode `\\12\catcode `\$12\catcode
  `\&12\catcode `\#12\catcode `\^12\catcode `\_12\catcode `\%12\relax}%
\providecommand \@@startlink[1]{}%
\providecommand \@@endlink[0]{}%
\providecommand \url  [0]{\begingroup\@sanitize@url \@url }%
\providecommand \@url [1]{\endgroup\@href {#1}{\urlprefix }}%
\providecommand \urlprefix  [0]{URL }%
\providecommand \Eprint [0]{\href }%
\providecommand \doibase [0]{http://dx.doi.org/}%
\providecommand \selectlanguage [0]{\@gobble}%
\providecommand \bibinfo  [0]{\@secondoftwo}%
\providecommand \bibfield  [0]{\@secondoftwo}%
\providecommand \translation [1]{[#1]}%
\providecommand \BibitemOpen [0]{}%
\providecommand \bibitemStop [0]{}%
\providecommand \bibitemNoStop [0]{.\EOS\space}%
\providecommand \EOS [0]{\spacefactor3000\relax}%
\providecommand \BibitemShut  [1]{\csname bibitem#1\endcsname}%
\let\auto@bib@innerbib\@empty
\bibitem [{\citenamefont {Shapiro}\ and\ \citenamefont {Brumer}(2012)}]{SB12}%
  \BibitemOpen
  \bibfield  {author} {\bibinfo {author} {\bibfnamefont {M.}~\bibnamefont
  {Shapiro}}\ and\ \bibinfo {author} {\bibfnamefont {P.}~\bibnamefont
  {Brumer}},\ }\href@noop {} {\emph {\bibinfo {title} {Quantum Control of
  Molecular Processes}}},\ \bibinfo {edition} {2nd}\ ed.\ (\bibinfo
  {publisher} {Wiley-VCH},\ \bibinfo {address} {Weinheim},\ \bibinfo {year}
  {2012})\BibitemShut {NoStop}%
\bibitem [{\citenamefont {Shu}\ and\ \citenamefont {Henriksen}(2011)}]{SH11}%
  \BibitemOpen
  \bibfield  {author} {\bibinfo {author} {\bibfnamefont {C.-C.}\ \bibnamefont
  {Shu}}\ and\ \bibinfo {author} {\bibfnamefont {N.~E.}\ \bibnamefont
  {Henriksen}},\ }\href {\doibase 10.1063/1.3582928} {\bibfield  {journal}
  {\bibinfo  {journal} {J. Chem. Phys.}\ }\textbf {\bibinfo {volume} {134}},\
  \bibinfo {eid} {164308} (\bibinfo {year} {2011})}\BibitemShut {NoStop}%
\bibitem [{\citenamefont {Shu}\ and\ \citenamefont {Henriksen}(2012)}]{SH12}%
  \BibitemOpen
  \bibfield  {author} {\bibinfo {author} {\bibfnamefont {C.-C.}\ \bibnamefont
  {Shu}}\ and\ \bibinfo {author} {\bibfnamefont {N.~E.}\ \bibnamefont
  {Henriksen}},\ }\href {\doibase 10.1063/1.3678013} {\bibfield  {journal}
  {\bibinfo  {journal} {J. Chem. Phys.}\ }\textbf {\bibinfo {volume} {136}},\
  \bibinfo {eid} {044303} (\bibinfo {year} {2012})}\BibitemShut {NoStop}%
\bibitem [{\citenamefont {Grinev}, \citenamefont {Shapiro},\ and\ \citenamefont
  {Brumer}(2013)}]{GSB13}%
  \BibitemOpen
  \bibfield  {author} {\bibinfo {author} {\bibfnamefont {T.}~\bibnamefont
  {Grinev}}, \bibinfo {author} {\bibfnamefont {M.}~\bibnamefont {Shapiro}}, \
  and\ \bibinfo {author} {\bibfnamefont {P.}~\bibnamefont {Brumer}},\ }\href
  {\doibase 10.1063/1.4775808} {\bibfield  {journal} {\bibinfo  {journal} {J.
  Chem. Phys.}\ }\textbf {\bibinfo {volume} {138}},\ \bibinfo {eid} {044306}
  (\bibinfo {year} {2013})}\BibitemShut {NoStop}%
\bibitem [{\citenamefont {Schlosshauer}(2007)}]{Sch07}%
  \BibitemOpen
  \bibfield  {author} {\bibinfo {author} {\bibfnamefont {M.}~\bibnamefont
  {Schlosshauer}},\ }\href@noop {} {\emph {\bibinfo {title} {Decoherence and
  the Quantum-To-Classical Transition}}}\ (\bibinfo  {publisher}
  {Springer-Verlag, Berlin},\ \bibinfo {year} {2007})\BibitemShut {NoStop}%
\bibitem [{\citenamefont {Elran}\ and\ \citenamefont {Brumer}(2013)}]{EB13}%
  \BibitemOpen
  \bibfield  {author} {\bibinfo {author} {\bibfnamefont {Y.}~\bibnamefont
  {Elran}}\ and\ \bibinfo {author} {\bibfnamefont {P.}~\bibnamefont {Brumer}},\
  }\href {\doibase 10.1063/1.4810009} {\bibfield  {journal} {\bibinfo
  {journal} {J. Chem. Phys.}\ }\textbf {\bibinfo {volume} {138}},\ \bibinfo
  {eid} {234308} (\bibinfo {year} {2013})}\BibitemShut {NoStop}%
\bibitem [{\citenamefont {Prokhorenko}\ \emph {et~al.}(2006)\citenamefont
  {Prokhorenko}, \citenamefont {Nagy}, \citenamefont {Waschuk}, \citenamefont
  {Brown}, \citenamefont {Birge},\ and\ \citenamefont {Miller}}]{PM&06}%
  \BibitemOpen
  \bibfield  {author} {\bibinfo {author} {\bibfnamefont {V.}~\bibnamefont
  {Prokhorenko}}, \bibinfo {author} {\bibfnamefont {A.}~\bibnamefont {Nagy}},
  \bibinfo {author} {\bibfnamefont {S.~A.}\ \bibnamefont {Waschuk}}, \bibinfo
  {author} {\bibfnamefont {L.~S.}\ \bibnamefont {Brown}}, \bibinfo {author}
  {\bibfnamefont {R.~R.}\ \bibnamefont {Birge}}, \ and\ \bibinfo {author}
  {\bibfnamefont {R.~J.~D.}\ \bibnamefont {Miller}},\ }\href {\doibase
  10.1126/science.1130747} {\bibfield  {journal} {\bibinfo  {journal}
  {Science}\ }\textbf {\bibinfo {volume} {313}},\ \bibinfo {pages} {1257}
  (\bibinfo {year} {2006})}\BibitemShut {NoStop}%
\bibitem [{\citenamefont {Joffre}(2007)}]{Jof07}%
  \BibitemOpen
  \bibfield  {author} {\bibinfo {author} {\bibfnamefont {M.}~\bibnamefont
  {Joffre}},\ }\href {\doibase 10.1126/science.1137011} {\bibfield  {journal}
  {\bibinfo  {journal} {Science}\ }\textbf {\bibinfo {volume} {317}},\ \bibinfo
  {pages} {453b} (\bibinfo {year} {2007})}\BibitemShut {NoStop}%
\bibitem [{\citenamefont {Prokhorenko}\ \emph {et~al.}(2007)\citenamefont
  {Prokhorenko}, \citenamefont {Nagy}, \citenamefont {Waschuk}, \citenamefont
  {Brown}, \citenamefont {Birge},\ and\ \citenamefont {Miller}}]{PM&07}%
  \BibitemOpen
  \bibfield  {author} {\bibinfo {author} {\bibfnamefont {V.}~\bibnamefont
  {Prokhorenko}}, \bibinfo {author} {\bibfnamefont {A.}~\bibnamefont {Nagy}},
  \bibinfo {author} {\bibfnamefont {S.~A.}\ \bibnamefont {Waschuk}}, \bibinfo
  {author} {\bibfnamefont {L.~S.}\ \bibnamefont {Brown}}, \bibinfo {author}
  {\bibfnamefont {R.~R.}\ \bibnamefont {Birge}}, \ and\ \bibinfo {author}
  {\bibfnamefont {R.~J.~D.}\ \bibnamefont {Miller}},\ }\href {\doibase
  10.1126/science.1137032} {\bibfield  {journal} {\bibinfo  {journal}
  {Science}\ }\textbf {\bibinfo {volume} {317}},\ \bibinfo {pages} {453c}
  (\bibinfo {year} {2007})}\BibitemShut {NoStop}%
\bibitem [{\citenamefont {Florean}\ \emph {et~al.}(2009)\citenamefont
  {Florean}, \citenamefont {Cardoza}, \citenamefont {White}, \citenamefont
  {Lanyi}, \citenamefont {Sension},\ and\ \citenamefont {Bucksbaum}}]{FC&09}%
  \BibitemOpen
  \bibfield  {author} {\bibinfo {author} {\bibfnamefont {C.}~\bibnamefont
  {Florean}}, \bibinfo {author} {\bibfnamefont {D.}~\bibnamefont {Cardoza}},
  \bibinfo {author} {\bibfnamefont {J.~L.}\ \bibnamefont {White}}, \bibinfo
  {author} {\bibfnamefont {J.~K.}\ \bibnamefont {Lanyi}}, \bibinfo {author}
  {\bibfnamefont {R.~J.}\ \bibnamefont {Sension}}, \ and\ \bibinfo {author}
  {\bibfnamefont {P.~H.}\ \bibnamefont {Bucksbaum}},\ }\href {\doibase
  10.1073/pnas.0904589106} {\bibfield  {journal} {\bibinfo  {journal} {Proc.
  Natl. Acad. Sci. U.S.A.}\ }\textbf {\bibinfo {volume} {106}},\ \bibinfo
  {pages} {10896} (\bibinfo {year} {2009})}\BibitemShut {NoStop}%
\bibitem [{\citenamefont {Katz}, \citenamefont {Ratner},\ and\ \citenamefont
  {Kosloff}(2010)}]{KRK10}%
  \BibitemOpen
  \bibfield  {author} {\bibinfo {author} {\bibfnamefont {G.}~\bibnamefont
  {Katz}}, \bibinfo {author} {\bibfnamefont {M.~A.}\ \bibnamefont {Ratner}}, \
  and\ \bibinfo {author} {\bibfnamefont {R.}~\bibnamefont {Kosloff}},\ }\href
  {http://stacks.iop.org/1367-2630/12/i=1/a=015003} {\bibfield  {journal}
  {\bibinfo  {journal} {New Journal of Physics}\ }\textbf {\bibinfo {volume}
  {12}},\ \bibinfo {pages} {015003} (\bibinfo {year} {2010})}\BibitemShut
  {NoStop}%
\bibitem [{\citenamefont {Spanner}, \citenamefont {Arango},\ and\ \citenamefont
  {Brumer}(2010)}]{SAB10}%
  \BibitemOpen
  \bibfield  {author} {\bibinfo {author} {\bibfnamefont {M.}~\bibnamefont
  {Spanner}}, \bibinfo {author} {\bibfnamefont {C.~A.}\ \bibnamefont {Arango}},
  \ and\ \bibinfo {author} {\bibfnamefont {P.}~\bibnamefont {Brumer}},\ }\href
  {\doibase 10.1063/1.3491366} {\bibfield  {journal} {\bibinfo  {journal} {J.
  Chem. Phys.}\ }\textbf {\bibinfo {volume} {133}},\ \bibinfo {pages} {151101}
  (\bibinfo {year} {2010})}\BibitemShut {NoStop}%
\bibitem [{\citenamefont {Kosloff}\ \emph {et~al.}(2011)\citenamefont
  {Kosloff}, \citenamefont {Ratner}, \citenamefont {Katz},\ and\ \citenamefont
  {Khasin}}]{KR&11}%
  \BibitemOpen
  \bibfield  {author} {\bibinfo {author} {\bibfnamefont {R.}~\bibnamefont
  {Kosloff}}, \bibinfo {author} {\bibfnamefont {M.}~\bibnamefont {Ratner}},
  \bibinfo {author} {\bibfnamefont {G.}~\bibnamefont {Katz}}, \ and\ \bibinfo
  {author} {\bibfnamefont {M.}~\bibnamefont {Khasin}},\ }\href {\doibase
  10.1016/j.proche.2011.08.040} {\bibfield  {journal} {\bibinfo  {journal}
  {Procedia Chemistry}\ }\textbf {\bibinfo {volume} {3}},\ \bibinfo {pages}
  {322 } (\bibinfo {year} {2011})},\ \bibinfo {note} {22nd Solvay Conference on
  Chemistry}\BibitemShut {NoStop}%
\bibitem [{\citenamefont {Prokhorenko}\ \emph {et~al.}(2011)\citenamefont
  {Prokhorenko}, \citenamefont {Halpin}, \citenamefont {Johnson}, \citenamefont
  {Miller},\ and\ \citenamefont {Brown}}]{PH&11}%
  \BibitemOpen
  \bibfield  {author} {\bibinfo {author} {\bibfnamefont {V.~I.}\ \bibnamefont
  {Prokhorenko}}, \bibinfo {author} {\bibfnamefont {A.}~\bibnamefont {Halpin}},
  \bibinfo {author} {\bibfnamefont {P.~J.~M.}\ \bibnamefont {Johnson}},
  \bibinfo {author} {\bibfnamefont {R.}~\bibnamefont {Miller}}, \ and\ \bibinfo
  {author} {\bibfnamefont {L.~S.}\ \bibnamefont {Brown}},\ }\href {\doibase
  10.1063/1.3554743} {\bibfield  {journal} {\bibinfo  {journal} {J. Chem.
  Phys.}\ }\textbf {\bibinfo {volume} {134}},\ \bibinfo {pages} {085105}
  (\bibinfo {year} {2011})}\BibitemShut {NoStop}%
\bibitem [{\citenamefont {{General Disucssion}}(2011)}]{faraday153}%
  \BibitemOpen
  \bibfield  {author} {\bibinfo {author} {\bibnamefont {{General
  Disucssion}}},\ }\href {\doibase 10.1039/C1FD90042K} {\bibfield  {journal}
  {\bibinfo  {journal} {Disc. Far. Soc.}\ }\textbf {\bibinfo {volume} {153}},\
  \bibinfo {pages} {428} (\bibinfo {year} {2011})}\BibitemShut {NoStop}%
\bibitem [{\citenamefont {Arango}\ and\ \citenamefont {Brumer}(2013)}]{AB13}%
  \BibitemOpen
  \bibfield  {author} {\bibinfo {author} {\bibfnamefont {C.~A.}\ \bibnamefont
  {Arango}}\ and\ \bibinfo {author} {\bibfnamefont {P.}~\bibnamefont
  {Brumer}},\ }\href {\doibase 10.1063/1.4792834} {\bibfield  {journal}
  {\bibinfo  {journal} {J. Chem. Phys.}\ }\textbf {\bibinfo {volume} {138}},\
  \bibinfo {eid} {071104} (\bibinfo {year} {2013})}\BibitemShut {NoStop}%
\bibitem [{\citenamefont {Pach\'on}, \citenamefont {Yu},\ and\ \citenamefont
  {Brumer}(2013)}]{PYB13}%
  \BibitemOpen
  \bibfield  {author} {\bibinfo {author} {\bibfnamefont {L.~A.}\ \bibnamefont
  {Pach\'on}}, \bibinfo {author} {\bibfnamefont {L.}~\bibnamefont {Yu}}, \ and\
  \bibinfo {author} {\bibfnamefont {P.}~\bibnamefont {Brumer}},\ }\href
  {\doibase 10.1039/C3FD20144A} {\bibfield  {journal} {\bibinfo  {journal}
  {Faraday Discussions}\ }\textbf {\bibinfo {volume} {163}},\ \bibinfo {pages}
  {485} (\bibinfo {year} {2013})},\ \Eprint {http://arxiv.org/abs/1212.6416}
  {arXiv:1212.6416} \BibitemShut {NoStop}%
\bibitem [{\citenamefont {Brumer}\ and\ \citenamefont {Shapiro}(1989)}]{BS89}%
  \BibitemOpen
  \bibfield  {author} {\bibinfo {author} {\bibfnamefont {P.}~\bibnamefont
  {Brumer}}\ and\ \bibinfo {author} {\bibfnamefont {M.}~\bibnamefont
  {Shapiro}},\ }\href {\doibase 10.1016/0301-0104(89)90013-X} {\bibfield
  {journal} {\bibinfo  {journal} {Chem. Phys.}\ }\textbf {\bibinfo {volume}
  {139}},\ \bibinfo {pages} {221} (\bibinfo {year} {1989})}\BibitemShut
  {NoStop}%
\bibitem [{\citenamefont {Pach\'on}, \citenamefont {Triana},\ and\
  \citenamefont {Brumer}(2013)}]{PTB13}%
  \BibitemOpen
  \bibfield  {author} {\bibinfo {author} {\bibfnamefont {L.~A.}\ \bibnamefont
  {Pach\'on}}, \bibinfo {author} {\bibfnamefont {J.~F.}\ \bibnamefont
  {Triana}}, \ and\ \bibinfo {author} {\bibfnamefont {P.}~\bibnamefont
  {Brumer}},\ }\href@noop {} {\bibfield  {journal} {\bibinfo  {journal}
  {Canonical Typicality Deviations at Low Temperature}\ ,\ \bibinfo {pages} {In
  preparation.}} (\bibinfo {year} {2013})}\BibitemShut {NoStop}%
\bibitem [{\citenamefont {Pach\'on}\ and\ \citenamefont
  {Brumer}(2013{\natexlab{a}})}]{PB13}%
  \BibitemOpen
  \bibfield  {author} {\bibinfo {author} {\bibfnamefont {L.~A.}\ \bibnamefont
  {Pach\'on}}\ and\ \bibinfo {author} {\bibfnamefont {P.}~\bibnamefont
  {Brumer}},\ }\href {\doibase 10.1103/PhysRevA.87.022106} {\bibfield
  {journal} {\bibinfo  {journal} {Phys. Rev. A}\ }\textbf {\bibinfo {volume}
  {87}},\ \bibinfo {pages} {022106} (\bibinfo {year} {2013}{\natexlab{a}})},\
  \Eprint {http://arxiv.org/abs/1210.6374} {arXiv:1210.6374} \BibitemShut
  {NoStop}%
\bibitem [{\citenamefont {Jiang}\ and\ \citenamefont {Brumer}(1991)}]{JB91}%
  \BibitemOpen
  \bibfield  {author} {\bibinfo {author} {\bibfnamefont {X.-P.}\ \bibnamefont
  {Jiang}}\ and\ \bibinfo {author} {\bibfnamefont {P.}~\bibnamefont {Brumer}},\
  }\href {\doibase 10.1063/1.460467} {\bibfield  {journal} {\bibinfo  {journal}
  {J. Chem. Phys.}\ }\textbf {\bibinfo {volume} {94}},\ \bibinfo {pages} {5833}
  (\bibinfo {year} {1991})}\BibitemShut {NoStop}%
\bibitem [{\citenamefont {Hoki}\ and\ \citenamefont {Brumer}(2011)}]{HB11}%
  \BibitemOpen
  \bibfield  {author} {\bibinfo {author} {\bibfnamefont {H.}~\bibnamefont
  {Hoki}}\ and\ \bibinfo {author} {\bibfnamefont {P.}~\bibnamefont {Brumer}},\
  }\href@noop {} {\bibfield  {journal} {\bibinfo  {journal} {Procedia Chem.}\
  }\textbf {\bibinfo {volume} {3}},\ \bibinfo {pages} {122} (\bibinfo {year}
  {2011})}\BibitemShut {NoStop}%
\bibitem [{\citenamefont {Brumer}\ and\ \citenamefont {Shapiro}(2012)}]{BS12b}%
  \BibitemOpen
  \bibfield  {author} {\bibinfo {author} {\bibfnamefont {P.}~\bibnamefont
  {Brumer}}\ and\ \bibinfo {author} {\bibfnamefont {M.}~\bibnamefont
  {Shapiro}},\ }\href {\doibase 10.1073/pnas.1211209109} {\bibfield  {journal}
  {\bibinfo  {journal} {Proc. Natl. Acad. Sci. U.S.A.}\ }\textbf {\bibinfo
  {volume} {109}},\ \bibinfo {pages} {19575} (\bibinfo {year}
  {2012})}\BibitemShut {NoStop}%
\bibitem [{\citenamefont {Pach\'on}\ and\ \citenamefont
  {Brumer}(2013{\natexlab{b}})}]{PB12b}%
  \BibitemOpen
  \bibfield  {author} {\bibinfo {author} {\bibfnamefont {L.~A.}\ \bibnamefont
  {Pach\'on}}\ and\ \bibinfo {author} {\bibfnamefont {P.}~\bibnamefont
  {Brumer}},\ }\href@noop {} {\bibfield  {journal} {\bibinfo  {journal} {J.
  Math. Phys. (submitted)}\ } (\bibinfo {year} {2013}{\natexlab{b}})},\ \Eprint
  {http://arxiv.org/abs/arXiv:1207.3104} {arXiv:arXiv:1207.3104} \BibitemShut
  {NoStop}%
\bibitem [{\citenamefont {Galve}, \citenamefont {Pach\'on},\ and\ \citenamefont
  {Zueco}(2010)}]{GPZ10}%
  \BibitemOpen
  \bibfield  {author} {\bibinfo {author} {\bibfnamefont {F.}~\bibnamefont
  {Galve}}, \bibinfo {author} {\bibfnamefont {L.~A.}\ \bibnamefont {Pach\'on}},
  \ and\ \bibinfo {author} {\bibfnamefont {D.}~\bibnamefont {Zueco}},\ }\href
  {\doibase 10.1103/PhysRevLett.105.180501} {\bibfield  {journal} {\bibinfo
  {journal} {Phys. Rev. Lett.}\ }\textbf {\bibinfo {volume} {105}},\ \bibinfo
  {pages} {180501} (\bibinfo {year} {2010})}\BibitemShut {NoStop}%
\bibitem [{\citenamefont {Feynman}\ and\ \citenamefont {Hibbs}(1965)}]{FH65}%
  \BibitemOpen
  \bibfield  {author} {\bibinfo {author} {\bibfnamefont {R.~P.}\ \bibnamefont
  {Feynman}}\ and\ \bibinfo {author} {\bibfnamefont {A.~R.}\ \bibnamefont
  {Hibbs}},\ }\href@noop {} {\emph {\bibinfo {title} {Quantum physics and path
  integrals}}}\ (\bibinfo  {publisher} {McGraw--Hill, New York},\ \bibinfo
  {year} {1965})\BibitemShut {NoStop}%
\bibitem [{\citenamefont {Ingold}(2002)}]{Ing02}%
  \BibitemOpen
  \bibfield  {author} {\bibinfo {author} {\bibfnamefont {G.-L.}\ \bibnamefont
  {Ingold}},\ }in\ \href {\doibase 10.1007/3-540-45855-7_1} {\emph {\bibinfo
  {booktitle} {Coherent Evolution in Noisy Environments}}},\ \bibinfo {series}
  {Lecture Notes in Physics}, Vol.\ \bibinfo {volume} {611},\ \bibinfo {editor}
  {edited by\ \bibinfo {editor} {\bibfnamefont {A.}~\bibnamefont
  {Buchleitner}}\ and\ \bibinfo {editor} {\bibfnamefont {K.}~\bibnamefont
  {Hornberger}}}\ (\bibinfo  {publisher} {Springer Berlin Heidelberg},\
  \bibinfo {year} {2002})\ pp.\ \bibinfo {pages} {1--53}\BibitemShut {NoStop}%
\bibitem [{\citenamefont {Pach\'on}, \citenamefont {Ingold},\ and\
  \citenamefont {Dittrich}(2010)}]{PID10}%
  \BibitemOpen
  \bibfield  {author} {\bibinfo {author} {\bibfnamefont {L.~A.}\ \bibnamefont
  {Pach\'on}}, \bibinfo {author} {\bibfnamefont {G.-L.}\ \bibnamefont
  {Ingold}}, \ and\ \bibinfo {author} {\bibfnamefont {T.}~\bibnamefont
  {Dittrich}},\ }\href {\doibase 10.1016/j.chemphys.2010.05.024} {\bibfield
  {journal} {\bibinfo  {journal} {Chem. Phys.}\ }\textbf {\bibinfo {volume}
  {375}},\ \bibinfo {pages} {209} (\bibinfo {year} {2010})}\BibitemShut
  {NoStop}%
\bibitem [{\citenamefont {Shapiro}\ and\ \citenamefont {Brumer}(2003)}]{MB03}%
  \BibitemOpen
  \bibfield  {author} {\bibinfo {author} {\bibfnamefont {M.}~\bibnamefont
  {Shapiro}}\ and\ \bibinfo {author} {\bibfnamefont {P.}~\bibnamefont
  {Brumer}},\ }\href {http://stacks.iop.org/0034-4885/66/i=6/a=201} {\bibfield
  {journal} {\bibinfo  {journal} {Rep. Prog. Phys.}\ }\textbf {\bibinfo
  {volume} {66}},\ \bibinfo {pages} {859} (\bibinfo {year} {2003})}\BibitemShut
  {NoStop}%
\bibitem [{\citenamefont {Feynman}\ and\ \citenamefont {Vernon}(1963)}]{FV63}%
  \BibitemOpen
  \bibfield  {author} {\bibinfo {author} {\bibfnamefont {R.~P.}\ \bibnamefont
  {Feynman}}\ and\ \bibinfo {author} {\bibfnamefont {F.~L.}\ \bibnamefont
  {Vernon}},\ }\href {\doibase 10.1016/0003-4916(63)90068-X} {\bibfield
  {journal} {\bibinfo  {journal} {Annals of Physics}\ }\textbf {\bibinfo
  {volume} {24}},\ \bibinfo {pages} {118} (\bibinfo {year} {1963})}\BibitemShut
  {NoStop}%
\bibitem [{\citenamefont {Caldeira}\ and\ \citenamefont
  {Leggett}(1983)}]{CL83}%
  \BibitemOpen
  \bibfield  {author} {\bibinfo {author} {\bibfnamefont {A.~O.}\ \bibnamefont
  {Caldeira}}\ and\ \bibinfo {author} {\bibfnamefont {A.~L.}\ \bibnamefont
  {Leggett}},\ }\href {\doibase 10.1016/0378-4371(83)90013-4} {\bibfield
  {journal} {\bibinfo  {journal} {Physica A}\ }\textbf {\bibinfo {volume}
  {121}},\ \bibinfo {pages} {587} (\bibinfo {year} {1983})}\BibitemShut
  {NoStop}%
\bibitem [{\citenamefont {Grabert}, \citenamefont {Schramm},\ and\
  \citenamefont {Ingold}(1988)}]{GSI88}%
  \BibitemOpen
  \bibfield  {author} {\bibinfo {author} {\bibfnamefont {H.}~\bibnamefont
  {Grabert}}, \bibinfo {author} {\bibfnamefont {P.}~\bibnamefont {Schramm}}, \
  and\ \bibinfo {author} {\bibfnamefont {G.-L.}\ \bibnamefont {Ingold}},\
  }\href {\doibase 10.1016/0370-1573(88)90023-3} {\bibfield  {journal}
  {\bibinfo  {journal} {Phys. Rep.}\ }\textbf {\bibinfo {volume} {168}},\
  \bibinfo {pages} {115} (\bibinfo {year} {1988})}\BibitemShut {NoStop}%
\bibitem [{\citenamefont {Weiss}(2012)}]{Wei12}%
  \BibitemOpen
  \bibfield  {author} {\bibinfo {author} {\bibfnamefont {U.}~\bibnamefont
  {Weiss}},\ }\href@noop {} {\emph {\bibinfo {title} {Quantum Dissipative
  Systems}}},\ \bibinfo {edition} {4th}\ ed.\ (\bibinfo  {publisher} {World
  Scientific, Singapore},\ \bibinfo {year} {2012})\BibitemShut {NoStop}%
\bibitem [{\citenamefont {Grabert}, \citenamefont {Weiss},\ and\ \citenamefont
  {Talkner}(1984)}]{GWT84}%
  \BibitemOpen
  \bibfield  {author} {\bibinfo {author} {\bibfnamefont {H.}~\bibnamefont
  {Grabert}}, \bibinfo {author} {\bibfnamefont {U.}~\bibnamefont {Weiss}}, \
  and\ \bibinfo {author} {\bibfnamefont {P.}~\bibnamefont {Talkner}},\ }\href
  {\doibase 10.1007/BF01307505} {\bibfield  {journal} {\bibinfo  {journal} {Z.
  Phys. B}\ }\textbf {\bibinfo {volume} {55}},\ \bibinfo {pages} {87} (\bibinfo
  {year} {1984})}\BibitemShut {NoStop}%
\bibitem [{\citenamefont {Haake}\ and\ \citenamefont {Reibold}(1985)}]{HR85}%
  \BibitemOpen
  \bibfield  {author} {\bibinfo {author} {\bibfnamefont {F.}~\bibnamefont
  {Haake}}\ and\ \bibinfo {author} {\bibfnamefont {R.}~\bibnamefont
  {Reibold}},\ }\href {\doibase 10.1103/PhysRevA.32.2462} {\bibfield  {journal}
  {\bibinfo  {journal} {Phys. Rev. A}\ }\textbf {\bibinfo {volume} {32}},\
  \bibinfo {pages} {2462} (\bibinfo {year} {1985})}\BibitemShut {NoStop}%
\bibitem [{\citenamefont {H\"{a}nggi}\ and\ \citenamefont
  {Ingold}(2005)}]{HI05}%
  \BibitemOpen
  \bibfield  {author} {\bibinfo {author} {\bibfnamefont {P.}~\bibnamefont
  {H\"{a}nggi}}\ and\ \bibinfo {author} {\bibfnamefont {G.-L.}\ \bibnamefont
  {Ingold}},\ }\href {\doibase 10.1063/1.1853631} {\bibfield  {journal}
  {\bibinfo  {journal} {Chaos}\ }\textbf {\bibinfo {volume} {15}},\ \bibinfo
  {pages} {026105} (\bibinfo {year} {2005})}\BibitemShut {NoStop}%
\bibitem [{\citenamefont {Campisi}, \citenamefont {Talkner},\ and\
  \citenamefont {H{\"a}nggi}(2009)}]{CTH09}%
  \BibitemOpen
  \bibfield  {author} {\bibinfo {author} {\bibfnamefont {M.}~\bibnamefont
  {Campisi}}, \bibinfo {author} {\bibfnamefont {P.}~\bibnamefont {Talkner}}, \
  and\ \bibinfo {author} {\bibfnamefont {P.}~\bibnamefont {H{\"a}nggi}},\
  }\href {\doibase 10.1103/PhysRevLett.102.210401} {\bibfield  {journal}
  {\bibinfo  {journal} {Phys. Rev. Lett.}\ }\textbf {\bibinfo {volume} {102}},\
  \bibinfo {pages} {210401} (\bibinfo {year} {2009})}\BibitemShut {NoStop}%
\bibitem [{Note1()}]{Note1}%
  \BibitemOpen
  \bibinfo {note} {A Mathematica 8.0 script with the numerical implementation
  of the results for the Ohmic spectral density can be found at \protect \href
  {http://gfam.udea.edu.co/~lpachon/scripts/} {http://gfam.udea.edu.co/$\sim
  $lpachon/scripts/oqsystems}.}\BibitemShut {Stop}%
\bibitem [{\citenamefont {H\"anggi}\ and\ \citenamefont {Jung}(2007)}]{HJ07}%
  \BibitemOpen
  \bibfield  {author} {\bibinfo {author} {\bibfnamefont {P.}~\bibnamefont
  {H\"anggi}}\ and\ \bibinfo {author} {\bibfnamefont {P.}~\bibnamefont
  {Jung}},\ }in\ \href {\doibase 10.1002/9780470141489.ch4} {\emph {\bibinfo
  {booktitle} {Advances in Chemical Physics}}},\ \bibinfo {series} {Lecture
  Notes in Physics}, Vol.\ \bibinfo {volume} {611},\ \bibinfo {editor} {edited
  by\ \bibinfo {editor} {\bibfnamefont {I.}~\bibnamefont {Prigogine}}\ and\
  \bibinfo {editor} {\bibfnamefont {S.~A.}\ \bibnamefont {Rice}}}\ (\bibinfo
  {publisher} {John Wiley \& Sons, Inc.},\ \bibinfo {year} {2007})\ pp.\
  \bibinfo {pages} {239--326}\BibitemShut {NoStop}%
\bibitem [{\citenamefont {Pach\'on}\ and\ \citenamefont
  {Brumer}(2013{\natexlab{c}})}]{PB13c}%
  \BibitemOpen
  \bibfield  {author} {\bibinfo {author} {\bibfnamefont {L.~A.}\ \bibnamefont
  {Pach\'on}}\ and\ \bibinfo {author} {\bibfnamefont {P.}~\bibnamefont
  {Brumer}},\ }\href@noop {} {\bibfield  {journal} {\bibinfo  {journal}
  {Experimentally Accessible Determination of Spectral Densities of Molecular
  Complexes}\ ,\ \bibinfo {pages} {In preparation}} (\bibinfo {year}
  {2013}{\natexlab{c}})}\BibitemShut {NoStop}%
\bibitem [{\citenamefont {Kast}\ and\ \citenamefont {Ankerhold}(2013)}]{KA13}%
  \BibitemOpen
  \bibfield  {author} {\bibinfo {author} {\bibfnamefont {D.}~\bibnamefont
  {Kast}}\ and\ \bibinfo {author} {\bibfnamefont {J.}~\bibnamefont
  {Ankerhold}},\ }\href {\doibase 10.1103/PhysRevLett.110.010402} {\bibfield
  {journal} {\bibinfo  {journal} {Phys. Rev. Lett.}\ }\textbf {\bibinfo
  {volume} {110}},\ \bibinfo {pages} {010402} (\bibinfo {year}
  {2013})}\BibitemShut {NoStop}%
\bibitem [{\citenamefont {Scholak}\ and\ \citenamefont {Brumer}(2013)}]{Sch13}%
  \BibitemOpen
  \bibfield  {author} {\bibinfo {author} {\bibfnamefont {T.}~\bibnamefont
  {Scholak}}\ and\ \bibinfo {author} {\bibfnamefont {P.}~\bibnamefont
  {Brumer}},\ }\href@noop {} {\bibfield  {journal} {\bibinfo  {journal} {Phys.
  Rev. Lett.}\ ,\ \bibinfo {pages} {submitted}} (\bibinfo {year} {2013})},\
  \Eprint {http://arxiv.org/abs/1305.4586} {arXiv:1305.4586} \BibitemShut
  {NoStop}%
\bibitem [{\citenamefont {Prokhorenko}, \citenamefont {Halpin},\ and\
  \citenamefont {Miller}(2011)}]{PHM11}%
  \BibitemOpen
  \bibfield  {author} {\bibinfo {author} {\bibfnamefont {V.~I.}\ \bibnamefont
  {Prokhorenko}}, \bibinfo {author} {\bibfnamefont {A.}~\bibnamefont {Halpin}},
  \ and\ \bibinfo {author} {\bibfnamefont {R.~J.~D.}\ \bibnamefont {Miller}},\
  }\href {\doibase 10.1039/C1FD00095K} {\bibfield  {journal} {\bibinfo
  {journal} {Faraday Discuss.}\ }\textbf {\bibinfo {volume} {153}},\ \bibinfo
  {pages} {27} (\bibinfo {year} {2011})}\BibitemShut {NoStop}%
\end{thebibliography}%

\appendix
\section{Evolution of the density operator}
\label{app:EvoDenOpe}

The matrix $\mathsf{A}$ contains the information about the open system
evolution and is defined as
\begin{align}
\label{equ:RotMatUniEvo}
\mathsf{A} &=
\frac{1}{q_0^6 \det \mathsf{M} }
\left(\begin{array}{cc}
1 + \mathsf{A}_{11}  & \mathsf{A}_{12}
\\
 \mathsf{A}_{12} & 1 + \mathsf{A}_{22}
\end{array}\right),
\end{align}
being
\begin{align}
\begin{split}
\mathsf{A}_{11} &=
\frac{m^2}{\hbar^2 \omega_0^2 \Lambda^2}\left(1 - 2\Lambda \omega_0 \right)
\left[\frac{\dot{G}_+(t)}{G_+(t)} S(t) - \dot{S}(t)\right]^2
+ \left(1 + 2\Lambda \omega_0\right) \frac{\dot{G}_+^2(t)}{\omega_0^2 G_+^2(t)}
+\frac{2 \Omega}{\omega_0},
\end{split}
\end{align}
\begin{align}
\begin{split}
\mathsf{A}_{12} &=
\frac{m^3}{\hbar^3\omega_0 \Lambda^2}
\frac{S(t)}{G_+(t)}
\left[\frac{\dot{G}_+(t)}{G_+(t)} S(t) - \dot{S}(t)\right]
\left[\frac{S(t)}{G_+(t)} - 2\Lambda \omega_0 \right],
\end{split}
\\
\begin{split}
\mathsf{A}_{22} &=
\frac{m^2}{\hbar^2 \omega_0^2 \Lambda^2}\left(1 - 2\Lambda \omega_0 \right)
\frac{S^2(t)}{G^2_+(t)}
+
\left(1 + 2\Lambda \omega_0\right) \frac{1}{\omega_0^2 G_+^2(t)}
-\frac{2 \Omega}{\omega_0}.
\end{split}
\end{align}
$\Lambda$ is related to the second moment in equilibrium of the $\hat{q}$
by means of $\langle \hat{q}^2\rangle = \hbar \Lambda / m$ while $\Omega$
through $\langle \hat{p}^2\rangle = \hbar m \Omega$ \cite{GSI88}.
$S(t)$ denotes the symmetric position autocorrelation function,
$S(t) = \frac{1}{2}\langle \hat{q}(t) \hat{q} + \hat{q} \hat{q}(t)\rangle$.
In the limit $\omega_{\mathrm{D}}\rightarrow \infty$, it is given by
\begin{align}
\begin{split}
\label{equ:Soft}
S(t) &= \frac{\hbar}{4m\omega_{\mathrm{d}}}
[\exp(-\lambda_2 t)\coth( \textrm{\textonehalf}\mathrm{i} \hbar \beta \lambda_2)
-
\exp(-\lambda_1 t) \coth(\textrm{\textonehalf} \mathrm{i} \hbar \beta \lambda_1)]
-\Gamma(t),
\end{split}
\end{align}
\begin{align}
\label{equ:gammaoft}
\Gamma(t) & = \frac{\gamma}{m \beta}\sum_{n=-\infty}^{\infty}
\frac{|\nu_n| \exp(-|\nu_n| t)}{(\omega_0^2 +\nu_n^2)^2 - \gamma^2 \nu_n^2},
\end{align}
being $\nu_n = 2\pi n /\hbar \beta$ the Matsubara frequencies (cf. Ref.~\citenum{GWT84,GSI88}).

The matrix $\mathsf{M}$ is given by
\begin{equation}
\mathsf{M}(t) = \frac{m}{\hbar}\left(
\begin{array}{cccc}
R^{--}(t) & -\mathrm{i} \frac{\dot{G}_+(t)}{G_+(t)}&
-C_2^-(t) + R^{+-}(t)&
\mathrm{i} \frac{1}{G_-(t)} - \mathrm{i} C_1^-(t)
\\
-\mathrm{i}\frac{\dot{G}_+(t)}{G_+(t)}& 0 & \mathrm{i} \frac{1}{G_+(t)} & 0
\\
-C_2^-(t) + R^{+-}(t) & \mathrm{i} \frac{1}{G_+(t)} &
- 2 C_2^+(t) + R^{++}(t)&
-\mathrm{i} \frac{\dot{G}_+(t)}{G_+(t)} - \mathrm{i} C_1^+(t)
\\
\mathrm{i} \frac{1}{G_-(t)} - \mathrm{i} C_1^-(t) & 0 &
-\mathrm{i} \frac{\dot{G}_+(t)}{G_+(t)} - \mathrm{i} C_1^+(t)& 0
\end{array}
\right),
\end{equation}
being
\begin{align}
C_j^+(t) =
\int\limits_0^t  \mathrm{d}s & C_j(s) \frac{G_+(t-s)}{G_+(t)},
\qquad
C_j^-(t) =
\int\limits_0^t \mathrm{d}s C_j(s) \frac{G_-(s)}{G_-(t)},
\end{align}
\begin{align}
R^{+-}(t) &=
\int_0^t \mathrm{d}s  \int_0^t \mathrm{d}u R(s,u) \frac{G_+(t-s)}{G_+(t)}
\frac{G_-(u)}{G_-(t)},
\end{align}
with $R^{++}(t)$ and $R^{--}(t)$ defined accordingly.

\end{document}